# The impacts of the electricity demand pattern on electricity system cost and the electricity supply mix: a comprehensive modeling analysis for Europe

Xiaoming Kan[a,*]   Lina Reichenberg[a,b]   Fredrik Hedenus[a]

[a]Department of Space, Earth and Environment, Chalmers University of Technology, Gothenburg, Sweden

[b]Department of Mathematics and Systems Analysis, Aalto University, Helsinki, Finland

**Abstract**

Energy system models for long-term planning are widely used to explore the future electricity system. Typically, to represent the future electricity demand in these models, historical demand profiles are used directly or scaled up linearly. Although the volume change for the electricity demand is considered, the potential change of the demand pattern is ignored. Meanwhile, the future electricity demand pattern is highly uncertain due to various factors, including climate change, e-mobility, electric heating, and electric cooling. We use a techno-economic cost optimization model to investigate a stylized case and assess the effects on system cost and electricity supply mix of assuming different demand patterns for the models. Our results show that differences in diurnal demand patterns affect the system cost by less than 3%. Similarly, demand profiles with a flat seasonal variation or a winter peak result in only minor changes in system cost, as compared to the present demand profile. Demand profiles with a summer peak may display a system cost increase of up to 8%, whereas the electricity supply mix may differ by a factor of two. A more detailed case study is conducted for Europe and the results are consistent with the findings from the stylized case.

## 1. Introduction

Energy system models for long-term investment planning are widely used to generate insights for policy analysis and decision making for the future electricity system. Such models typically minimize total system costs under technologic, environmental and policy constraints and generate a cost-effective electricity system portfolio. Often, perfect foresight with regard to the future technology costs, resource availability, and electricity demand is assumed. However, these models have been criticized for their inability to capture the uncertainties surrounding the future electricity system [1-4]. With large uncertainties that grow over time, singular projections in energy system models often fall short of the full spectrum of the plausible future electricity system and produce misleading results. Currently, there is a growing body of studies addressing the uncertainties associated with technological, economic, and societal parameters. These studies relate to future technology investment costs [5], weather conditions [6, 7], discount rates [8], policies [9], and indeed, demand growth [10, 11]. Many studies focusing on the future electricity demand have either directly used the historical demand profile or have linearly scaled up the historical demand profile to a new value as the future electricity demand [12-16]. Although the volume of annual electricity demand is considered, the inter-temporal pattern of the electricity demand profile (demand pattern) is assumed to remain the same.

* Email: kanx@chalmers.se

It is well established in the literature that both the volume of annual electricity demand and the demand pattern are heavily influenced by factors such as population expansion, economic growth, climate change, e-mobility, electric heating (EH), electric cooling (EC), and technological innovations [10, 11, 17-21]. Specifically, some recent studies have highlighted the impacts of electric vehicles (EVs), EH and EC on the demand pattern. Boßmann and Staffell [10] estimated that the peak demand in the UK would increase by 50% due to the extensive diffusion of EVs and EH by 2050. A slightly milder, but still large, increase of 28% for the peak demand due to the wide adoption of EVs and EH in the UK was found by Pudjianto et al. [19] if demand-side management (DSM) is implemented. Staffell and Pfenninger [11] estimated a 20% increase in peak demand around 2030 in the UK due to EH only. Kannan [20] found that the summer peak in Switzerland would increase by 2%–23% in 2050 due to increased use of air conditioners (ACs). These studies underline the substantial uncertainties regarding the future electricity demand pattern.

Some studies have estimated that radical changes in the electricity demand pattern might strongly affect the electricity system [10, 11, 19]. The impacts might be more evident for a renewable electricity system as it is less capable of load following due to the intermittency of variable renewable energy (VRE) resources, as compared with the conventional electricity system based on dispatchable thermal power plants [22]. Therefore, given the high level of uncertainty regarding the future electricity demand pattern, it is important to understand how the changes in the electricity demand profile affect the modeling results, particularly in terms of the electricity supply mix and the system cost for the renewable electricity system. Several studies [4, 23, 24] have already evaluated the impacts of the volume change for the annual electricity demand on the electricity system, and found that a higher electricity demand might lead to more investment in renewable energy and a higher electricity cost. Even though several analyses have suggested that the electricity demand pattern may change rather dramatically in the coming decades, only two studies [24, 25] have explicitly analyzed the impacts of changes in the demand pattern on the electricity system. Zappa et al. [24] evaluated the impacts of different diurnal variations of the electricity demand on the level of investment in a renewable European electricity system using an optimization approach. They found that a higher diurnal variation leads to a slightly higher level of electricity generation due to more curtailment, whereas the impact on system cost is minor. Likewise, Boßmann et al. [25] analyzed the change in diurnal variation of the electricity demand and showed that a higher diurnal variation results in more investment in dispatchable generation capacities to deal with the peak demand, and slightly increases the electricity generation.

Some other studies have adopted cost optimization models to evaluate the impacts of DSM on the system cost and the composition of the electricity system with high penetration of renewables. Behboodi et al. [26] discovered that different DSM-related flexibilities (0%–10% of the hourly electricity demand) exert only a minor influence on the system cost. Similarly, Domínguez and Carrión [27] showed that DSM has weak impacts on investments in generation capacity and on the system cost, though it has a noticeable effect on system operation. Taljegard et al. [28] assessed EVs as a variation management strategy for the electricity system and reported that optimized charging of EVs slightly reduces the system cost compared to direct charging based on the owners' driving patterns.

The literature on DSM thus suggests that the shifting and curtailment of demand have limited impacts on electricity system cost, and this may be due to a limited potential (only a minor share of the load is shifted and curtailed in a short temporal duration) or high implementation costs. Still, future changes in the electricity demand patterns are unlikely to be limited to minor changes in the diurnal variation but may also entail large changes in diurnal variation (for example due to EV charging) and seasonal variation (for example due to EH and EC). It remains unknown as to how these potential changes in diurnal and seasonal demand patterns might affect the cost and supply mix of the future electricity system. Therefore, it is not clear whether or not the practice of using (scaled-up) historical electricity demand profiles in energy system models produces potentially unacceptable errors regarding the system cost and electricity supply mix. In the present study, we fill this gap in the knowledge and evaluate the conditions under which the demand pattern is important for the modeling results. Specifically, we address the following question: *What is the effect on the system cost and the electricity supply mix from applying different demand patterns in energy system models?*

By investigating this question, we aim at providing insights regarding whether or not the changes in the future demand pattern can be disregarded for energy system modeling practice. To resolve this question, we use techno-economic cost optimization models for the electricity system to investigate a stylized case[1] involving three regions in Europe and one full-scale applied case (Europe). The paper is organized as follows. The model and input data are introduced in Section 2. In Section 3, the modeling results are presented. The mechanisms behind the results are then discussed in Section 4, and conclusions are drawn in Section 5. The model-specific code, input data, and output data will be made available online to ensure the transparency and reproducibility of the results.

**2. Methods**

---

[1] The stylized case refers to a stylized set-up for the electricity demand profiles. It has several different scenarios depending on the specific shape of the electricity demand profiles.

In the first stage of the present study, a stylized case that involves three regions with VRE resource endowments typical for Europe (Fig. 1) is investigated with a simple cost optimization model. The three regions are located in the south, central and north of Europe respectively and they are named as: South, Central and North. Region South is provided with data for VRE resources and electricity demand pattern from Spain plus Portugal. Similarly, data for VRE resources and electricity demand patterns for Germany and Norway are assigned to region Central and North respectively. The reason why we investigate a simplified stylized case is to look at regions with typical VRE resource potentials, to explore numerous possible demand patterns while keeping a reasonable computation time, and to make the results easy to analyze. Region South has good solar resources and region North is characterized by good wind resources. Region Central displays both solar- and wind resources, yet neither as good as those in South nor North, see Fig. 2. Since our main purpose is to analyze the impact of the demand pattern on an interconnected electricity system, there is the option to invest in transmission connections between the three nodes. The interrelated electricity system in the stylized case is modeled for one year with an hourly time resolution, given a cap on $CO_2$ emission expressed in grams of $CO_2$ per kWh of electricity demand. The effects of different demand patterns on the system cost and electricity supply mix are analyzed. In order to validate the results obtained from the stylized case, the REX model [29] is used to evaluate the European electricity system, by comparing scenarios with different demand patterns.

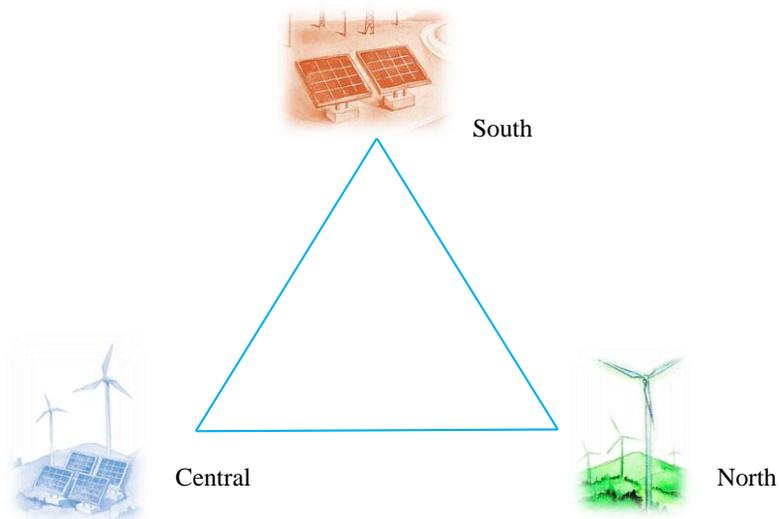

Fig. 1. Regions covered in the stylized case. We select three regions with typical VRE resource potentials and connect them with transmission grids to analyze the impacts of the demand pattern on an interconnected electricity system.

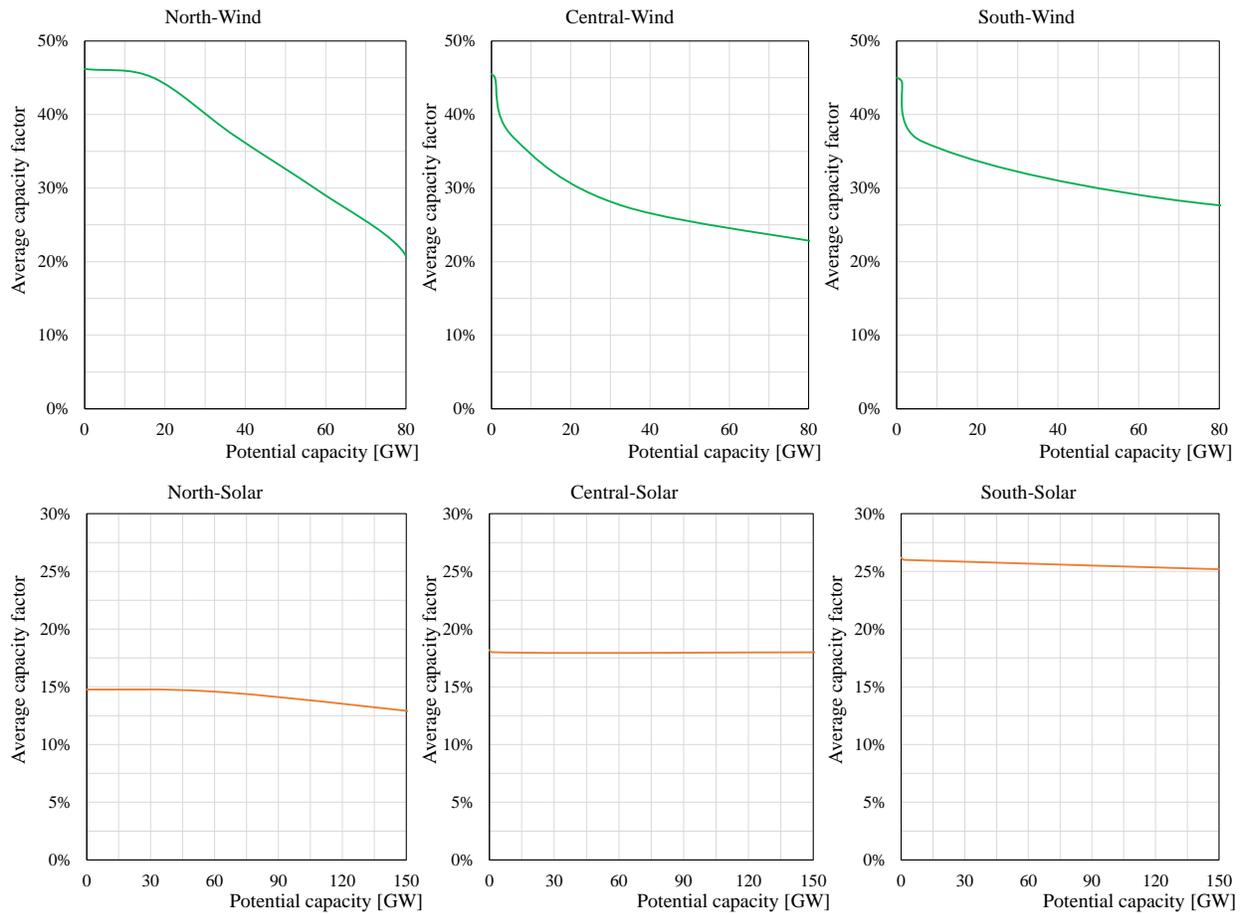

Fig. 2. Resource endowments for wind and solar in each region for the stylized case. The average capacity factor refers to the average hourly capacity factor for wind and solar in one year. All the data are collected through the GIS model developed by Mattsson et al. [30].

**2.1 The energy system model for the stylized case**

The model developed for the stylized case is a greenfield cost optimization model for capacity investments and the dispatch of electricity generation, transmission and storage. It employs an overnight investment approach to identify the minimum cost portfolio for the future electricity system. This entails a linear optimization problem with the objective to minimize the total annual electricity system cost, given the constraints of meeting the electricity demand, the renewable energy resource potentials, and a $CO_2$ emission cap. An overview of the model, the generation technology options, and the variation management strategies are depicted in Fig. 3.

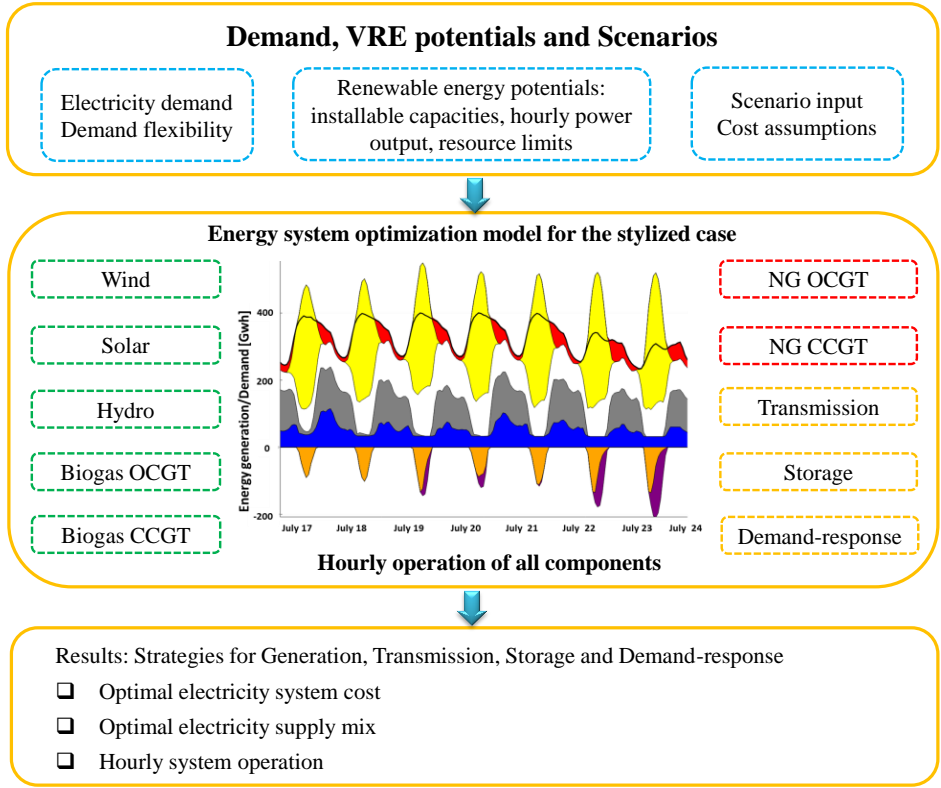

Fig. 3. Overview of the model for the stylized case. OCGT refers to open-cycle gas turbine; CCGT refers to combined-cycle gas turbine; NG refers to natural gas. The input data are shown in the blue dashed box. The renewable generation technologies are listed in the green dashed box. The fossil fuel fired generation technologies are presented in the red dashed box and variation management strategies are included in the orange dashed box.

The nodes in the model are labeled by $r$, $n$ represents the electricity generation technology at the node, $m$ represents the demand-response at the node, and $t$ is the time of the year. The total annual system cost consists of fixed annualized costs $C_n$ for electricity generation capacity $G_{rn}$, fixed annualized costs $C^{storage}$ for storage $S_r$, fixed annualized costs $C_{rr'}$ for transmission capacity $Z_{rr'}$, variable costs $R_n$ for electricity generation $g_{rnt}$ and variable costs $R_m$ for demand-response $d_{rmt}$. For storage and transmission, the variable cost is assumed to be zero. Therefore, the objective function of this linear optimization problem is formulated as follows:

$$Min \sum_{r,n} C_n G_{rn} + \sum_{r} C^{storage} S_r + \sum_{r,r'} 0.5 C_{rr'} Z_{rr'} + \sum_{r,n,t} R_n g_{rnt} + \sum_{r,m,t} R_m d_{rmt}. \tag{1}$$

Since $Z_{rr'}$ and $Z_{r'r}$ represent the capacity for the same transmission line $rr'$, a coefficient of 0.5 is incorporated into the transmission cost formula to avoid double counting.

The electricity demand has to be satisfied through generation, demand-response, trade and storage.

$$\sum_n g_{rnt} + \sum_m d_{rmt} + \sum_{r'} (\eta_\gamma \gamma_{r'rt} - \gamma_{rr't}) + (\eta_s \alpha_{rt} - \beta_{rt}) \geq D_{rt}, \tag{2}$$

where $g_{rnt}$ is the electricity generation, $d_{rmt}$ is the demand-response, $\gamma_{rr't}$ is the electricity traded from node $r$ to node $r'$, $\eta_\gamma$ is the efficiency of transmission, $\alpha_{rt}$ is the discharge from storage, $\beta_{rt}$ is the charge into storage, $\eta_s$ is the round-trip efficiency of storage and $D_{rt}$ is the hourly electricity demand.

For the other constraints imposed on the optimization problem and a more detailed description of the model, please see Appendix B in the supplementary material.

## 2.2 Demand data and scenarios for the stylized case

The electricity demand data for the three regions are taken from ENTSO-E [31], and they are scaled down to the same value 87 TWh (the average electricity demand in 2014 for European countries). We adopt a lower electricity demand for each region because in a renewable electricity system, scarcity of supply may occur when there is a high electricity demand and restricted land availability for VRE, and this strongly influences the modeling results [32]. A lower electricity demand can mitigate the impacts of land use constraints on the modeling results and can reveal more clearly the impacts of different electricity demand patterns. The demand profiles are then manipulated so as to display typical, stylized seasonal and diurnal variations (see Fig. 4). We do not make a detailed estimation of all the possible future demand patterns. Still, the stylized demand patterns created in this study do represent the typical possible future electricity demand profiles with regard to the potentials for EV charging, EH, EC, etc. By applying these typical demand profiles as input to the model and contrasting the modeling results, we are then able to evaluate the magnitude of error in modeling results if the historical demand profile is scaled up as input to the model. The stylized seasonal variations for the demand profile consist of the following six types:

1. Current demand pattern (N), whereby the demand profile is maintained in its current shape (Fig. 4a–c);
2. Zero seasonal variation (A), such that there is no seasonal variation for the demand profile (Fig. 4d–f);
3. Medium winter peak (W), whereby the annual peak demand (the maximum electricity demand in 1 year) is in the wintertime and the seasonal variation (the maximum peak demand in winter minus the minimum peak demand in summer) is 20% of the annual peak demand (Fig. 4g);
4. High winter peak (W+), whereby the annual peak demand is in the wintertime and the seasonal variation is 40% of the annual peak demand (Fig. 4h);
5. Medium summer peak (S), such that the annual peak demand is in the summertime and the seasonal variation (the maximum peak demand in summer minus the minimum peak demand in winter) is 20% of the annual peak demand (Fig. 4i); and
6. High summer peak (S+), whereby the annual peak demand is in the summertime and the seasonal variation is 40% of the annual peak demand (Fig. 4j).

There are three types of stylized diurnal variations for the demand profile:

1. Zero diurnal variation (Zero) (Fig. 4d);

2. Medium diurnal variation (Medium), such that the diurnal variation (the maximum electricity demand minus the minimum electricity demand in 1 day) is 25% of the daily peak demand (maximum electricity demand in 1 day) (Fig. 4e); and

3. High diurnal variation (High), whereby the diurnal variation is 45% of the daily peak demand (Fig. 4f).

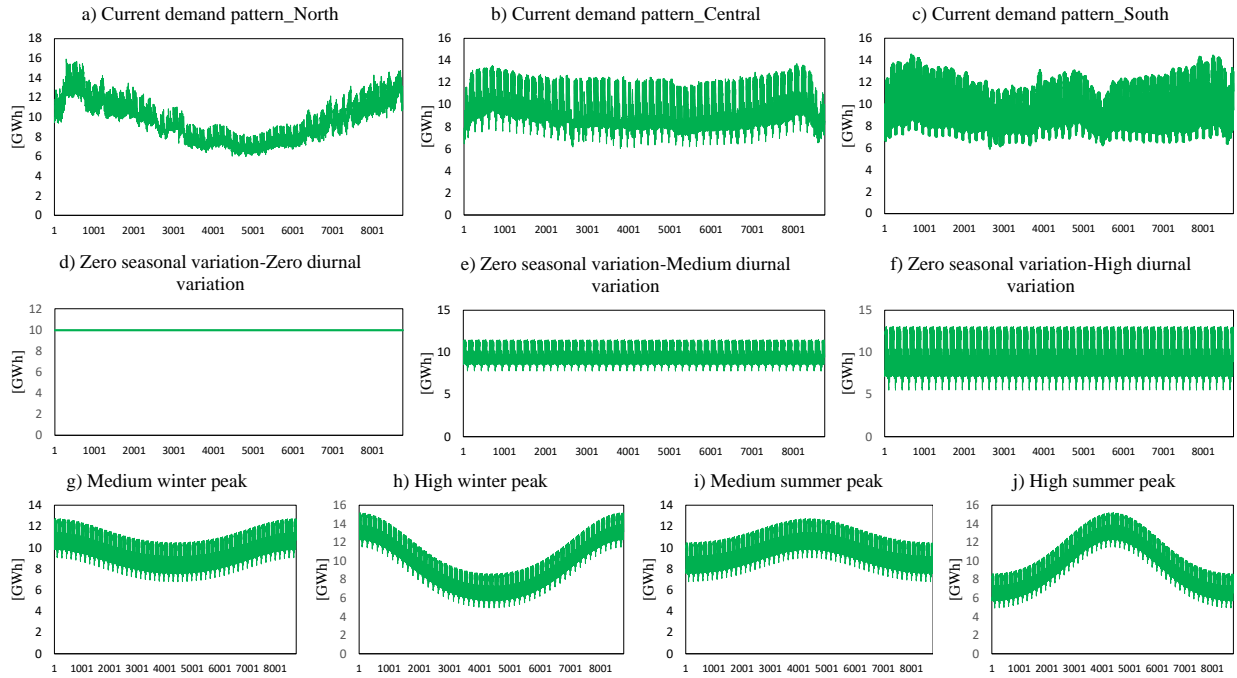

Fig. 4. Different electricity demand patterns in the stylized case.

Considering the climate situation in each region and that the seasonal variation is mainly due to heating/cooling, the summer peak (S, S+) is excluded for region North, the high summer peak (S+) is excluded for region Central, and the high winter peak (W+) is excluded for region South. All the other demand patterns for each region are regarded as the possible future demand patterns for these three regions and they are shown in Fig. 5. The demand data input to the optimization model are combinations of seasonal and diurnal variations of the electricity demand profiles for the three regions. In total there are 145 combinations. These combinations are categorized according to the shape of the aggregated demand profile[2] into six Groups labeled *Current demand pattern*, *Zero seasonal variation*, *Medium winter peak*, *High winter peak*, *Medium summer peak*, and *High summer peak*. Each combination is one Scenario and there could be several different combinations (Scenarios) in each Group. For the Group of the *Current demand pattern*, there is only one Scenario, with the three regions North, Central and South maintaining the current demand pattern (NNN). The Scenario of the *Current demand pattern* is the base Scenario for this study. The

---

[2] Under each combination of seasonal and diurnal variations of the three regional demand profiles, if we sum the hourly electricity demand for the three regions, we get an aggregated demand profile. We categorize the different combinations based on the shape of the aggregated demand profile. The aggregated demand profile is used as a benchmark to define scenarios for the present study. It is not the input demand data for the model. The three regional demand profiles are the input to the model.

Scenarios in the other five Groups display three different diurnal variations (*Zero*, *Medium*, *High*) depending on the shape of the aggregated demand profile. An overview of the Scenarios for the stylized case is given in Table 1.

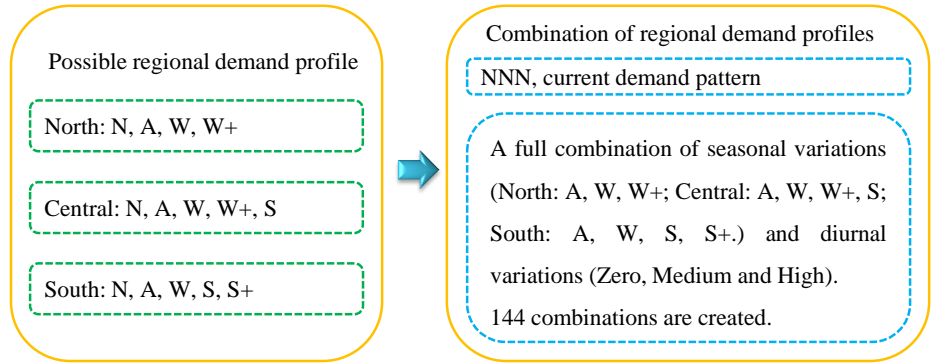

Fig. 5. Demand profile for each region and the combinations of the three regional demand profiles. The possible electricity demand profiles for each region are presented in the green dashed box and the combinations are shown in the blue dashed box.

**Table 1** Scenarios for the stylized case.

| Scenario[a] | Seasonal Variation | Diurnal variation | Detailed combination[b] (Sequence: North, Central, South) |
|---|---|---|---|
| Current demand pattern | Current pattern | Current pattern | NNN |
| Zero seasonal variation | Zero | Zero/Medium/High | AAA, ASW, AWS, WAS, WSA, AW+S+, W+AS+ |
| Medium winter peak | <20% of annual peak demand, winter peak | Zero/Medium/High | W+SS, AAW, AWA, WAA, WWS, WSW, WW+S+, W+WS+, AW+S, W+AS, W+SA, AWW, WWA, WAW, AW+A, W+AA, WW+S, W+WS, W+SW, W+W+S+ |
| High winter peak | ≥20% of annual peak demand, winter peak | Zero/Medium/High | WWW, WW+A, AW+W, W+AW, W+WA, W+W+S, W+W+A, WW+W, W+WW, W+W+W |
| Medium summer peak | <14% of annual peak demand, summer peak | Zero/Medium/High | WWS+, ASA, AAS, WSS, W+SS+, AWS+, WAS+ |
| High summer peak | ≥14% of annual peak demand, summer peak | Zero/Medium/High | ASS, AAS+, WSS+, ASS+ |

[a]Each Scenario represents a combination of seasonal and diurnal variations of the demand profiles for the three regions. All the combinations are categorized into six Scenario Groups based on the shape of the aggregated demand profile. Different combinations might result in a similar shape for the aggregated demand profile. Therefore, inside each Scenario Group, there are several different detailed combinations.

[b]For each detailed combination other than NNN, the three regional demand profiles have the same diurnal variation and the amplitude of the diurnal variation has three levels: *Zero*, *Medium* and *High*.

## 2.3 Input data for the stylized case

In this study, the transmission grids are assumed to be high-voltage direct current (HVDC) connections and the trade in electricity is represented as a simple transport problem [15, 16]. The length of the transmission line is measured as the distance between the population center of each region [30]. All the sub-regions in the model are treated as "copper plates" without intra-regional transmission constraints. The cost for the battery is used as a reference for storage in the model. However, the storage option could be any other storage technology with a similar cost structure.

The capacities of reservoir hydropower (hydro reservoir) and run-of-river hydropower (hydro RoR) are based on a previous report [33], and, to match the down scaling of the electricity demand in each region, the values are scaled down to 2.2 GW and 1.4 GW respectively. The hydro inflow is taken from reference [15] and scaled down so that the contribution of hydropower is 10% of the annual electricity demand in each region. In order to consider the downstream ecosystem and human needs for water, the minimum environmental flow [34, 35] of a hydro reservoir is set to 5% of the mean annual inflow. The fuel supply for biogas power plants is limited to a maximum of 5% of the annual electricity consumption, which is approximately the annual level of production of biogas from manure, agricultural residues and waste. Demand-response is one of the variation management options. Specifically, the aggregated consumers can curtail up to 5% of the demand at fixed costs within a given time period. More detailed description of the demand-response is provided in Table B.1 in the supplementary material.

The assumptions made as to wind and solar photovoltaic (PV) densities (W/m$^2$) and available land are listed in Table 2. The available land is given as a share of the suitable land, which is equivalent to the total land excluding populated areas, natural parks, lakes, mountains, etc. The input data for VRE are obtained with the GIS model developed by Mattsson et al. [30]. To represent more accurately the capacity factors for wind and solar power, the wind and solar technologies are divided into five classes based on resource quality. The wind speed is translated into capacity factors based on the power curve for a typical wind farm equipped with Vestas 112 3.075 MW wind turbines. Solar irradiation is used to calculate the capacity factor profiles with the assumption that the PV technology is fixed-latitude-tilted. The data for calculating the capacity factors for wind and solar power generation are obtained from the Global Wind Atlas [36] and the ECMWF ERA5 database [37]. All the data for VRE profiles are based on the values for Year 2018.

**Table 2** Assumptions made as to capacity limits for wind and solar photovoltaic. The density is the power output per unit area of a typical solar or wind farm.

|  | Solar Photovoltaic | Wind Onshore | Wind Offshore |
|---|---|---|---|
| Density [W/m$^2$] | 45 | 5 | 8 |
| Available land [%] | 6% | 10% | 33% |

The carbon cap is 10 g/kWh, which is roughly equivalent to a 98% reduction in $CO_2$ emission for the European electricity sector, as compared with the level of emission in Year 1990. The emission factor for natural gas is 198 $gCO_2$/kWh heat. The cost data and technical parameters for the main technologies are summarized in Table 3. These data are based on the cost projections for Year 2050 and are mainly taken from a previous report [38]. The initial investment cost is converted to the annualized cost with a discount rate of 5%.

**Table 3** Cost data and technical parameters.

| Technology | Investment cost [$/kW] | Variable O&M costs [$/MWh] | Fixed O&M costs [$/kW/yr] | Fuel costs [$/MWh fuel] | Lifetime [years] | Efficiency/ Round-trip efficiency |
|---|---|---|---|---|---|---|
| Natural gas OCGT | 493[a] | 3.6 | 12.5 | 36[b] | 30 | 0.35 |
| Natural gas CCGT | 800 | 3.6 | 10.5 | 36[b] | 30 | 0.6 |
| Biogas OCGT | 493[a] | 3.6 | 12.5 | 36[b] | 30 | 0.35 |
| Biogas CCGT | 800 | 3.6 | 10.5 | 36[b] | 30 | 0.6 |
| Onshore wind | 997 | 0 | 33 | n/a | 25 | n/a |
| Offshore wind | 2805[a] | 0 | 93 | n/a | 25 | n/a |
| Solar | 674 | 0 | 8 | n/a | 25 | n/a |
| Hydro reservoir | 2464[a] | 0 | 25 | n/a | 80 | n/a |
| Hydro RoR | 3696[a] | 0 | 74 | n/a | 80 | n/a |
| Transmission[c] | 479 $/MWkm | 0 | 9.6 $/MWkm | n/a | 40 | 0.016 loss per 1000 km |
| Converter[c] | 180 | 0 | 3.6 | n/a | 40 | 0.986 |
| Battery[d] | 156 $/kWh | 0 | 0 | n/a | 10 | 0.9 |

[a]Schröder et al. [39]

[b]Eurostat [40]

[c]Hagspiel et al. [41]

[d]Cole et al. [42]

To investigate further the breadth of conditions under which the choice of the demand pattern affects the modeling results, we conduct sensitivity analyses for the stylized case using different costs for wind, solar, transmission, and storage. Three levels of costs are assigned to each of the four technologies: "Low", "Medium", and "High". The detailed cost assumptions are listed in Table B.2 in the supplementary material. The reasons why the costs for wind, solar, transmission and storage are selected for the sensitivity analyses over all the other input parameters are: 1) these parameters are assumed to be the most important for the development of a VRE-based system; and 2) rather large uncertainties are attributed to these costs [43-46]. We also increase the carbon cap to 50 g/kWh to assess how the availability of flexible generation capacity affects the impact of the demand pattern on modeling results.

**2.4 The case of Europe**

The more detailed REX model [29], which is an energy system model used for policy support, is adopted to investigate the case of Europe. The countries included are the EU-28 (excluding Cyprus and Malta) plus Switzerland, Norway, Serbia, Bosnia and Herzegovina, North Macedonia, and Montenegro. These countries are divided into 13 regions and all the regions are assumed to be connected with HVDC grids, see Fig. 6. The synthetic electricity demand for Europe is created using the method of Mattsson et al. [30] for Year 2050. The demand profiles are treated in a manner similar to that used for the stylized case, so as to represent different seasonal variations and diurnal variations for the demand profile. The Scenarios for the case of Europe are presented in Table 4.

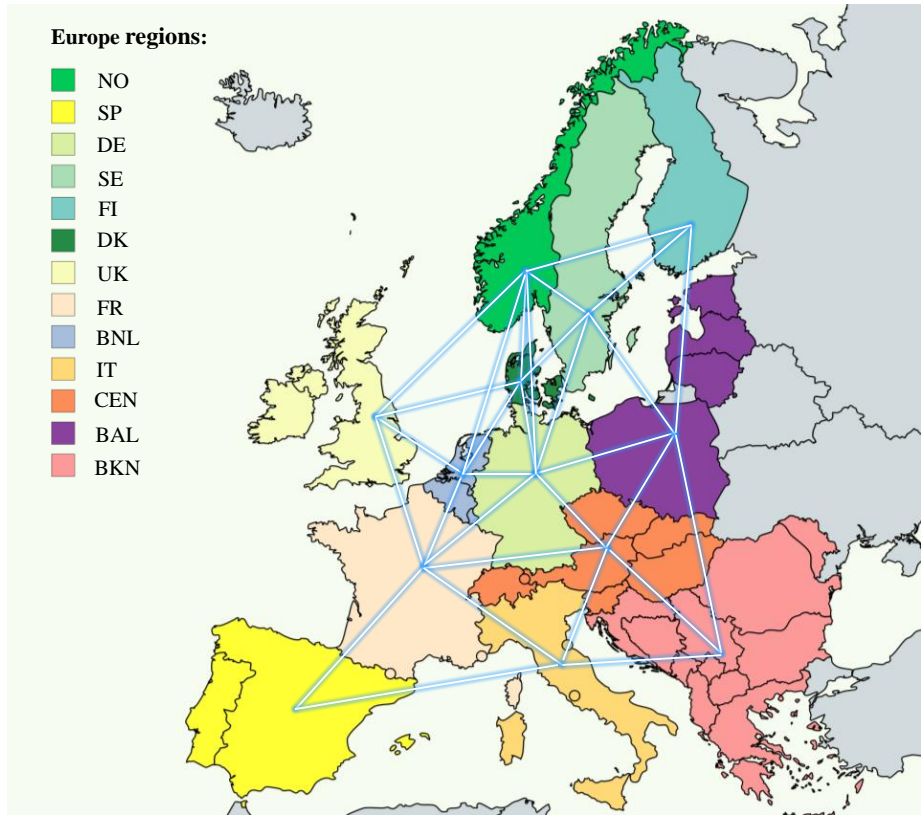

Fig. 6. Modeled regions and the interconnected transmission networks for Europe.

**Table 4** Scenarios for the case of Europe.

| Scenario[a] | Seasonal variation | Diurnal variation |
|---|---|---|
| Current demand pattern | Current pattern | Current pattern |
| Zero seasonal variation | Zero | Medium/High |
| Medium winter peak | 15% of annual peak demand, winter peak | Medium/High |
| High winter peak | 35% of annual peak demand, winter peak | Medium/High |
| Medium summer peak | 18% of annual peak demand, summer peak | Medium/High |
| High summer peak | 24% of annual peak demand, summer peak | Medium/High |

[a]More detailed description of the scenarios for the case of Europe is presented in Table B.3 in the supplementary material.

As for the input data for the case of Europe, the capacities of hydro reservoir and hydro RoR are kept constant at the current level due to environmental regulations in force. The capacity of hydropower is taken from the ENTSO-E statistics [33]. The inflow for each country is based on a previous study [15], and this value is divided into reservoir and RoR inflow based on the share of installed hydropower capacity. The assumptions made for the biogas power plants, wind, solar, transmission, storage, demand-response, carbon cap, technology cost, and discount rate are the same as those in the stylized case. The model is then run for three weather years with different wind outputs for Europe. The main results are calculated based on the year 2005 when the wind output is at the average level [32]. 2008 and 2010 are selected for sensitivity analysis due to their higher (2008) and lower (2010) wind output [32]. By calculating the results for different weather years, we aim to investigate how the variation in output for wind, solar and hydropower on an interannual basis may affect the impacts of demand patterns on the modeling results.

To further evaluate whether the conclusions drawn from the stylized case hold for more detailed modeling analyses used for policy support, the electricity system cost and electricity supply mix for the case of Europe are compared with the results obtained from the stylized case. A summary of the method for this study is presented in Fig. 7.

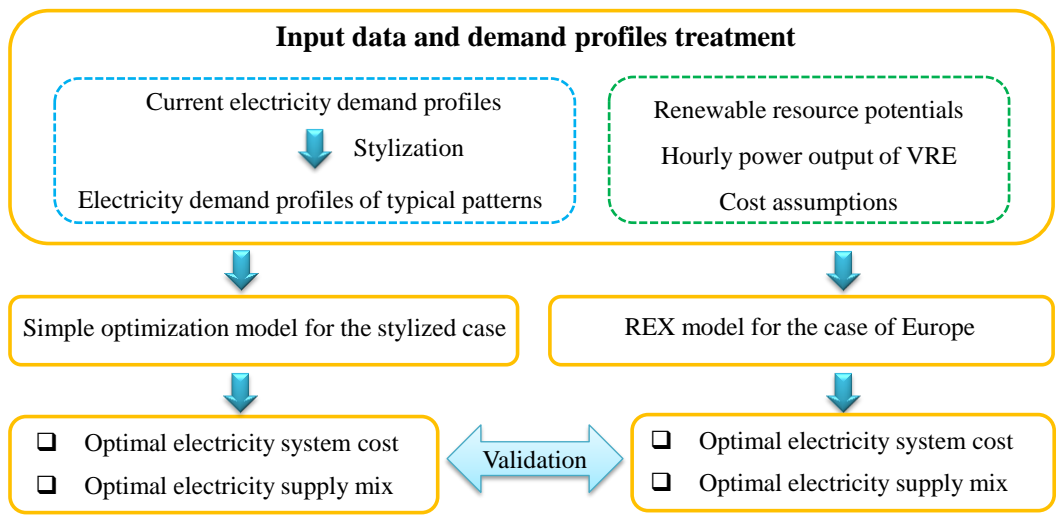

Fig. 7. Overview of the method. The input data for electricity demand and the stylization of demand profiles are shown in the blue dashed box. All the other input data are listed in the green dashed box.

## 3. Results

In this section, we present 1) the electricity system cost increases and the deviations in the electricity supply mix for the Scenarios of different seasonal variations compared with the Scenario of the *Current demand pattern*, and 2) the differences in electricity system cost and the deviations in the electricity supply mix for the Scenario of *High diurnal variation* compared with the Scenario of *Medium diurnal variation*. It is then possible to understand how different demand patterns affect the modeling results, particularly with respect to electricity system cost and the electricity supply mix. The mechanisms behind the impact of different demand patterns are further explained in Section 4.

### 3.1 Impacts of different seasonal variations for the stylized case

Fig. 8 shows how different seasonal variations of the demand profile affect the electricity system cost. Specifically, the figure illustrates how the average electricity system cost increases for Scenarios of different seasonal variations, as compared to the Scenario of the *Current demand pattern*. All the results are obtained under the conditions of medium diurnal variation, optimal transmission connection, and a carbon cap of 10 g/kWh. It is evident that if the annual peak of the electricity demand is in the winter (possibly due to large-scale deployment of electric heating), the system cost increase is low (<2% for all the Scenarios). Similarly, there is a low increase in the cost if the demand profile has no seasonal variation. In comparison, the system cost increases more if the annual peak is in the summer (possibly due to massive adoption of ACs). In such a case, the system cost increase is in the range of 3%–8%, depending on the amplitude of the summer peak.

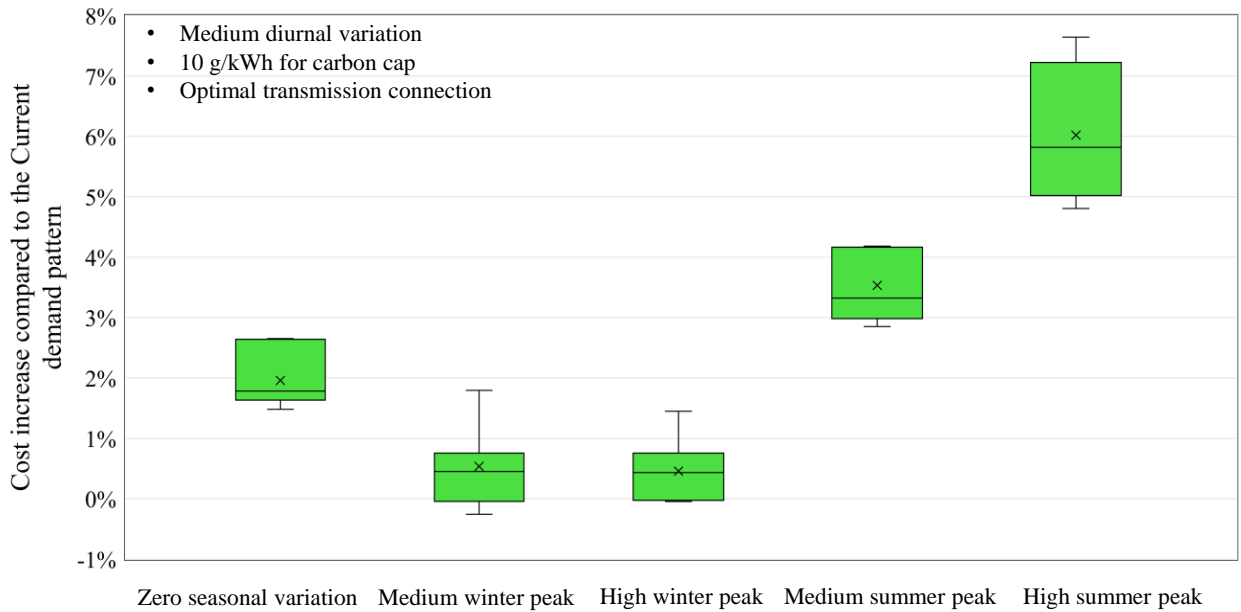

Fig. 8. Increases in the average electricity system cost for the Scenarios of different seasonal variations, as compared to the Scenario of the *Current demand pattern*. Each seasonal demand pattern (label on the *x*-axis) represents a Group of Scenarios with the same or similar aggregated demand profiles (for more details, see Table 1). The ends of the box are the upper and lower quartiles, so the box spans the interquartile range. The bar in the box represents the median value and the cross represents the average value. The whiskers are the two lines outside the box that extend to the highest and lowest values.

The optimal generation and storage capacity mixes under the Scenarios of different seasonal variations are shown in Fig. 9. We choose one typical Scenario for each seasonal demand pattern to simplify the figure. The other members of the Scenario Group (see Table 1) display similar changes in capacity mix. For Scenarios with zero seasonal variation and a winter peak, the electricity supply mix is similar to that for the current demand pattern. In contrast, scenarios with a summer peak display larger solar and storage capacity and are therefore quite distinct from the capacity mix that arises from using the current demand pattern.

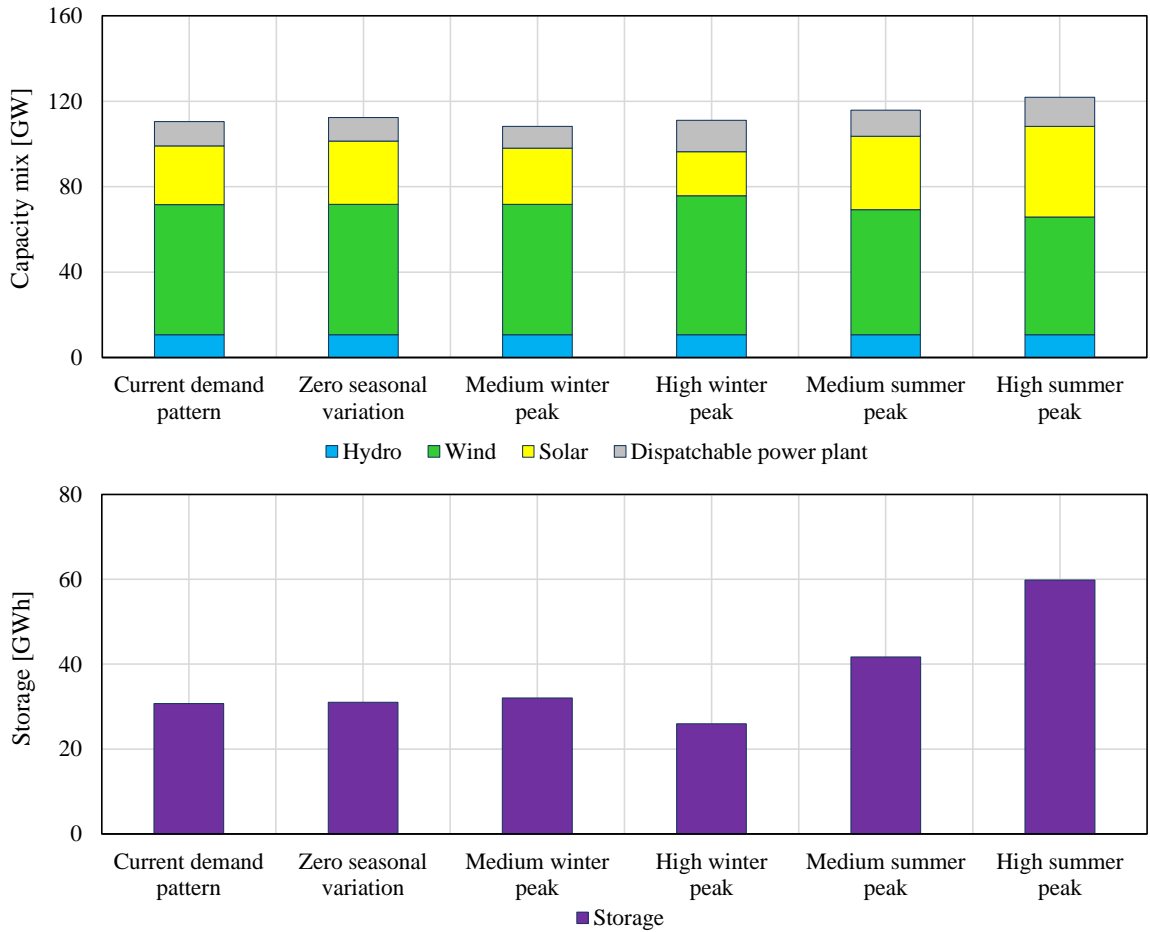

Fig. 9. Installed generation and storage capacities under the Scenarios of different seasonal variations. The capacity mix and storage for each seasonal demand pattern presented in this figure show the results of a sample scenario. All the members of each Scenario Group (see Table 1) have similar changes in capacity mix.

A summary of the relationship between the difference in system cost and the deviations in the electricity supply mix for the Scenarios of different seasonal variations are shown in Fig. 10. The overall change in system cost is estimated as a maximum of 8%. In stark contrast, the deviation in the electricity supply mix is much larger. In the Scenario with the highest summer peak, the increase in system cost is 8%, while the investments in solar and storage capacities increase by 54% and 95%, respectively, as compared to the Scenario of the *Current demand pattern*. Similar phenomena are observed for other Scenarios. Therefore, it is clear that a change in seasonal demand pattern has a stronger impact on the electricity supply mix than on the system cost.

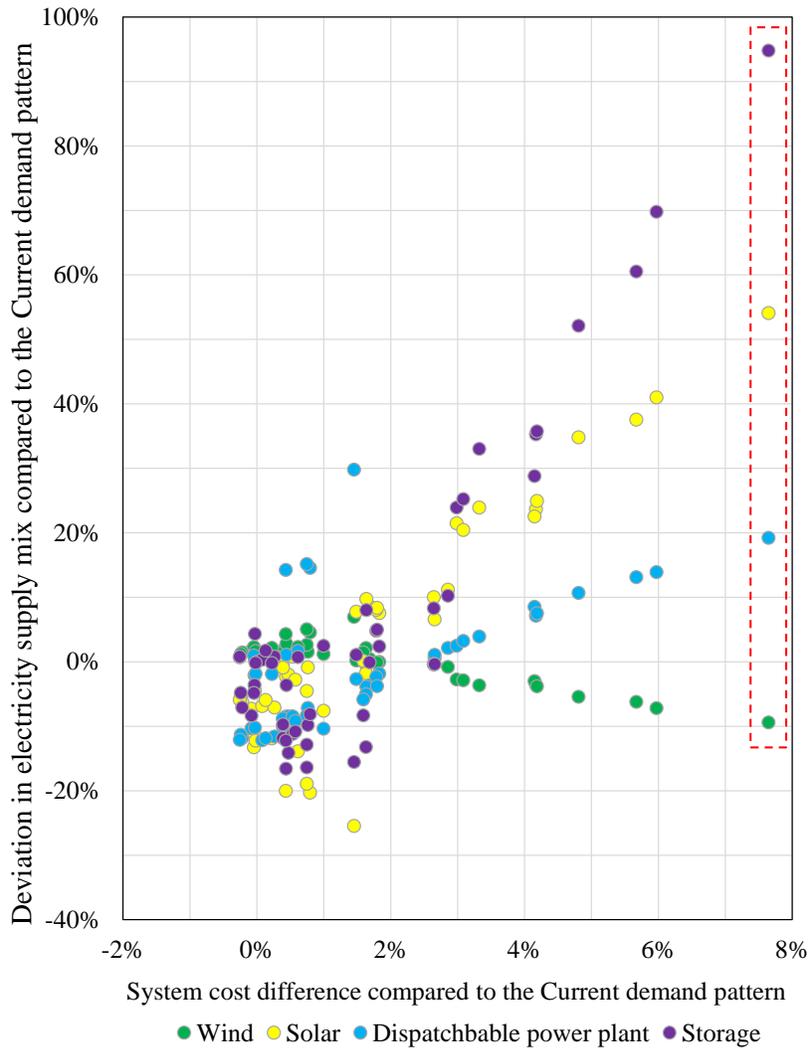

Fig. 10. The relationship between the difference in system cost and the deviations in the electricity supply mix for the Scenarios of different seasonal variations, as compared to the Scenario of the *Current demand pattern*. The dots inside the red rectangle represent the Scenario with the highest summer peak, as described in the text.

**3.2 Impacts of different diurnal variations for the stylized case**

The impacts of different diurnal variations (the cause of which may be various charging strategies for EVs) of the demand profile on the electricity system cost are depicted in Fig. 11. The figure shows the difference in electricity system cost for the Scenario of *High diurnal variation* compared to the Scenario of *Medium diurnal variation* under conditions of different seasonal demand patterns. Note first that a higher diurnal variation slightly increases the system cost regardless of the seasonal demand pattern. The impact of a higher diurnal variation is more evident for the demand profiles with a winter peak, while its influence is minor for the demand profiles with zero seasonal variation and a summer peak. In general, the difference in system cost between the Scenarios of *Medium-* and *High diurnal variation* is limited (<3%).

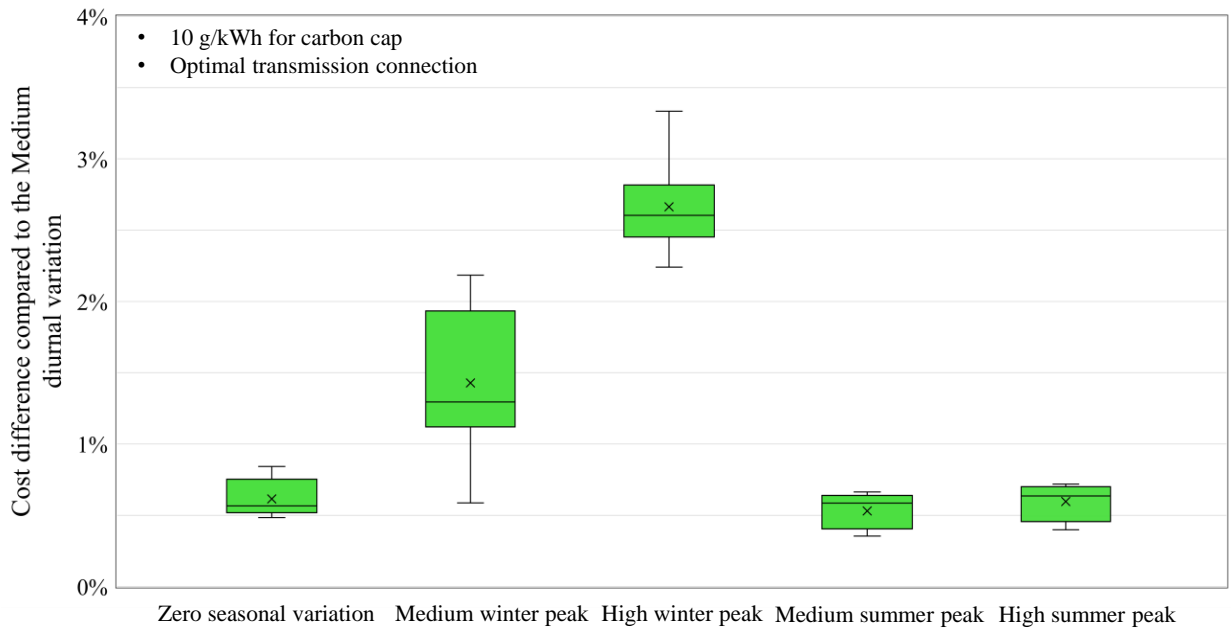

Fig. 11. Difference in the average electricity system cost for the Scenario of *High diurnal variation* compared to the Scenario of *Medium diurnal variation* under conditions of different seasonal demand patterns. Each seasonal demand pattern (label on the *x*-axis) represents a Group of Scenarios with the same or similar aggregated demand profiles (for more details, see Table 1). The ends of the box are the upper and lower quartiles, so the box spans the interquartile range. The bar in the box represents the median value and the cross represents the average value. The whiskers are the two lines outside the box that extend to the highest and lowest values.

Fig. 12 shows the relationship between the difference in system cost and the deviations in the electricity supply mix for the Scenario of *High diurnal variation* compared to the Scenario of *Medium diurnal variation*. Compared with the limited difference in the cost, there are substantial changes in the electricity supply mix, especially with regard to the capacities for solar and storage, in the Scenarios with a higher diurnal variation. In most cases, the system cost is higher for the Scenario with zero diurnal variation than for the Scenario with a diurnal variation (Fig. A.1). This may seem counterintuitive, since the experience of a power system based on thermal power plants may have instilled in us the notion that any demand variation precipitates a need for peaking plants. As these plants have lower utilization times and higher running costs than base-load plants, they entail a higher system cost. Yet, here we show that it no longer holds for a renewable electricity system with a large share of VRE.

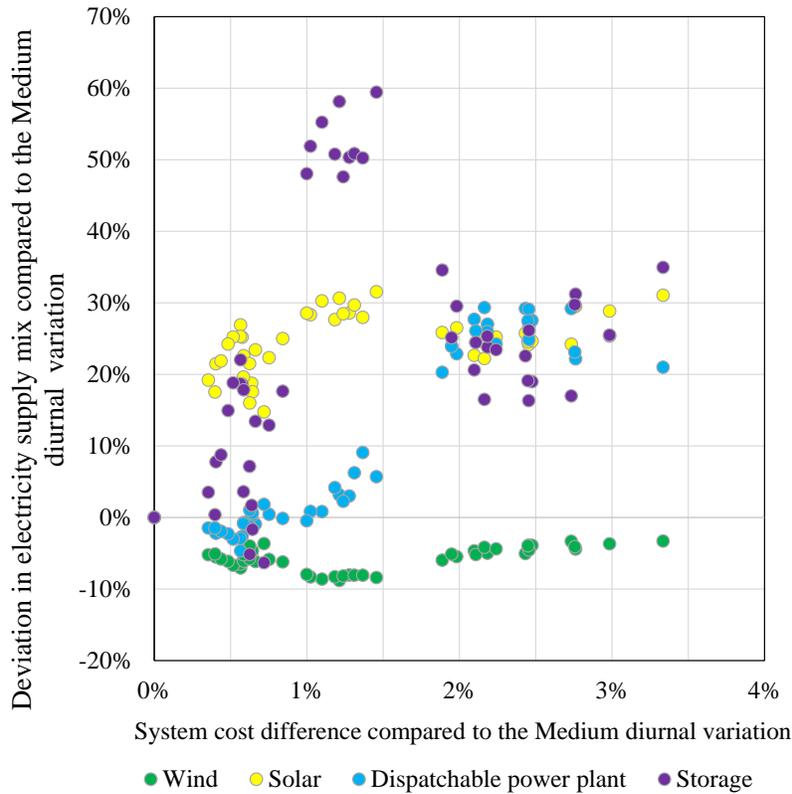

Fig. 12. The relationship between the difference in the system cost and the deviations in the electricity supply mix for the Scenario of *High diurnal variation* compared to the Scenario of *Medium diurnal variation*.

### 3.3 Impacts of flexible generation capacity for the stylized case

To further understand how the availability of flexible generation capacity affects the impact of the demand pattern on modeling results, we have also investigated a case with a higher carbon cap (50 g/kWh). A more generous carbon cap reduces the system cost increase due to a different seasonal variation for the electricity demand profile (Fig. A.2). This is mainly because a less-stringent carbon cap allows for more generation capacity and more energy from dispatchable natural gas power plants. In such a case, the electricity system is better able to follow the change in the electricity demand profile, which means that a different demand pattern has a smaller impact on the electricity system and the corresponding system cost. A similar phenomenon would occur if we were to allow additional dispatchable generation technologies, such as hydropower (Fig. A.2), fossil fuel plus CCS (carbon capture and storage) and nuclear power, in the modeling.

### 3.4 Sensitivity analysis

In the stylized case, the system cost increase is minor (<3%) for the Scenarios with zero seasonal variation and a winter peak, while the cost increases by up to 8% for a summer peak. We also conducted sensitivity analyses with different costs for wind, solar, transmission and storage, to see how the costs of the key technologies would affect the modeling results resulting from different demand patterns. The sensitivity analyses were conducted for one typical Scenario of the six different Scenario groups: *Current demand pattern*, *Zero seasonal variation*, *Medium winter peak*, *High winter peak*, *Medium summer peak*, and *High summer peak*. This choice ensures a wide coverage of cost differences resulting from different seasonal demand patterns and entails a reasonable overall computation time.

Fig. 13 shows how wind, solar, transmission and storage costs affect the differences in system cost between Scenarios with different seasonal variations and the Scenario of the *Current demand pattern*. Regardless of the cost parameters, the deviations in system cost due to different seasonal variations are consistent, with a greater increase in system cost for the summer peak than for the winter peak. Specifically, a low cost for solar power diminishes the system cost increase (to 5%) for the summer peak, and a high solar cost drives this value up to 9%. Similarly, a low cost for storage abates the system cost increase (to 3%) for the summer peak. A high cost for wind reduces the system cost increase for the summer peak, while the cost of transmission has little impact on the system cost.

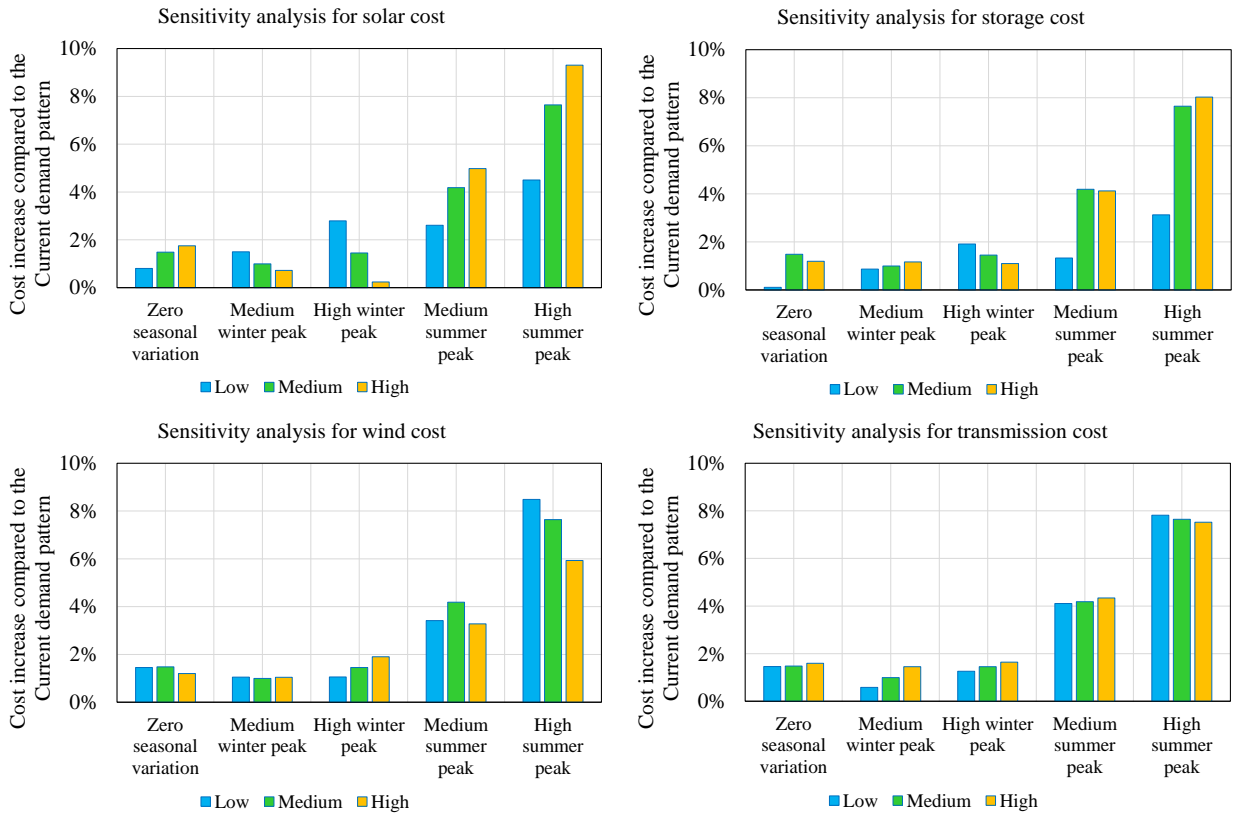

Fig. 13. Increases in the average electricity system cost for the Scenarios of different seasonal variations compared to the Scenario of the *Current demand pattern* under different cost assumptions for wind, solar, transmission and storage.

### 3.5 Modeling results for the case of Europe

To understand whether the results from the stylized case hold for a full-scale model, we ran the REX model for Europe, to analyze how different demand patterns affect the modeling results for an energy system model used for policy support. As is shown in Fig. 14, there is a greater increase in system cost for the summer peak (up to 8%) than for the winter peak (<2%), as compared with the Scenario of the *Current demand pattern*. This result holds true when varying the weather years as input to the model, see Fig. A.3 in the supplementary material. The variation in output for wind, solar and hydropower on an interannual basis has no significant influence on the impact of different seasonal demand patterns on system cost. The deviations in system cost due to different seasonal demand patterns for Europe as a whole are consistent with the results obtained for the stylized case, where the cost increase is less than 2% for the winter peak and up to 8% for the summer peak. Similar to the stylized case, a higher diurnal variation has only a minor impact on the system cost for Europe (Fig. A.4) and the change in the demand pattern has a greater impact on the electricity supply mix than on the system cost (Figs. A.5, A.6).

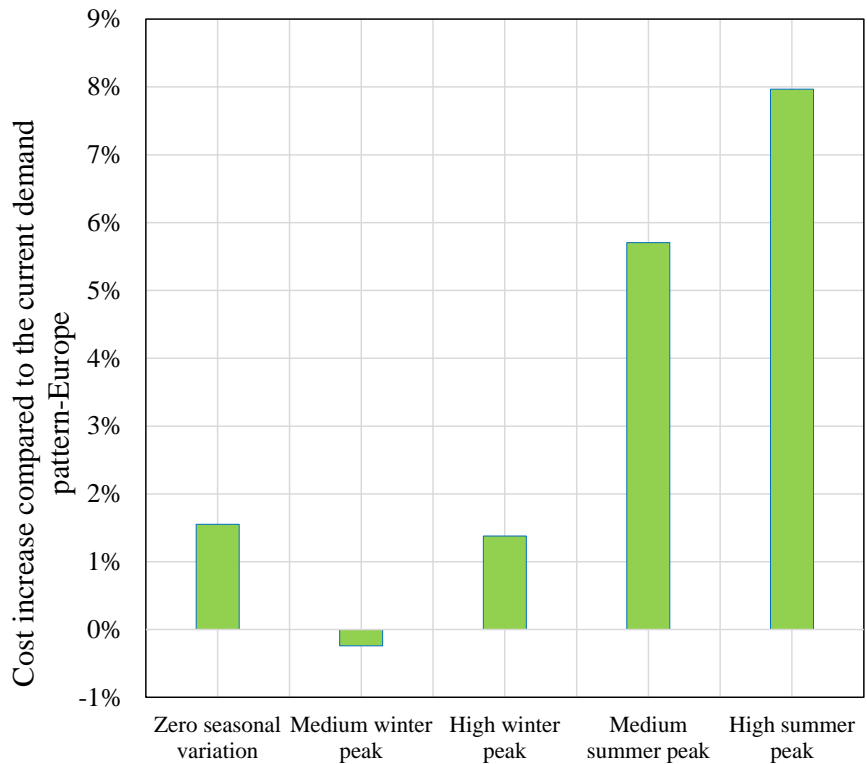

Fig. 14. Differences in the average electricity system cost between the Scenarios of different seasonal variations and the Scenario of the *Current demand pattern* for Europe. The input data for wind, solar and hydropower are based on the Year 2005.

## 4. Discussion

Through investigating the impacts of different seasonal demand patterns on the system cost, we find that a summer peak can increase the system cost by up to 8%, while the impacts of a winter peak and zero seasonal variation are limited (<3% increase in cost). This is because, in the stylized case, onshore wind power is cheap to install, and wind power has a typical seasonality with higher output in the wintertime than in the summertime. In addition, the variation of large-scale wind power can be smoothed through the expansion of transmission grids. Therefore, when the annual peak of the electricity demand is in winter, the seasonal variation of the demand profile is in line with the seasonal pattern of wind power, and cheap wind resource is deployed. In contrast, if the annual peak demand is in summer when the output of wind power is lower, the optimal system configuration contains more solar power and storage, which drives up the system cost. Correspondingly, there are large deviations in the capacity mix for the optimal electricity system portfolio, especially with respect to the solar and storage capacities. Overall, the deviations in the electricity supply mix are more evident than the changes in the system cost. The IEA report [47] analyzed the impact of increased cooling demand (a higher summer demand) on the electricity system and showed that a higher cooling demand increases the electricity cost as well as the total generation capacity. In the present study, both the electricity cost and the total generation capacity increase for the Scenarios with a summer peak as compared to the Scenario of the *Current demand pattern*, which is consistent with the conclusion of [47]. Similarly, Zhu et al. [48] assessed the effect of increased summer demand on the European electricity system, and estimated that a higher summer demand results in more installment of solar PV in South Europe, but the electricity cost remains stable. The results of Zhu et al. [48] are consistent with our findings regarding the impact of a higher summer peak on the electricity supply mix. The impact of a higher summer peak on electricity cost reported by Zhu et al. [48] is less influential than what we show, possibly due to the low amplitude of the summer peak in their electricity demand profile.

In reality, the electricity demand pattern is evolving over time due to climate change and increased sector coupling. The summer in Europe is becoming hotter, which will lead to the adoption of more ACs for cooling and, correspondingly, a higher demand in the summertime, as well as possibly a more pronounced daily peak. This is a possible case in which the demand pattern is genuinely influential. In such a case, if the modeler uses the historical electricity demand profile or linearly scales it up as the future electricity demand, misleading results might be produced, especially regarding the electricity capacity mix.

The impacts of different seasonal demand patterns on system cost are consistent, with a greater cost increase for a summer peak than for a winter peak, regardless of the cost assumptions for wind, solar, storage and transmission. As expected, a lower cost for solar and storage reduces the cost increase for the summer peak. This is because a lower cost for solar and storage avoids the system cost escalation that results from the large increases in solar and storage capacities. In addition, a more generous carbon cap or additional dispatchable generation capacity enables better load following for the system, which reduces the cost increase due to different seasonal demand patterns.

As for diurnal variation, the overall impact of different diurnal variations on the electricity system cost is limited (less than 3% increase in cost). Zappa et al. [24] analyzed investments in the European renewable electricity system under scenarios of different diurnal variations and found that a higher diurnal variation results in slightly more electricity generation, but the system cost is essentially unchanged. Similarly, Taljegard et al. [28] found that the optimal charging strategy for EVs has a minor impact on the system cost, as compared to direct charging based on the owners' driving patterns. The findings reported previously [24, 28] are consistent with our results on the impacts of different diurnal demand patterns on the electricity system cost. In the present study, more dispatchable generation capacities are installed for Europe if the demand profile displays a higher diurnal variation, which is in line with the main findings in reference [25].

As for the case of Europe, the system cost increase for Scenarios with zero seasonal variation and a winter peak is <2% while the summer peak increases the system cost by up to 8%. The cost deviation due to different seasonal demand patterns for Europe is in line with the results from the stylized case. Therefore, our results regarding the impacts of different seasonal demand patterns based on the stylized case are valid for Europe. These results might not be universally true for regions with different resource endowments, see Appendix C for a contrasting example.

Note that we do not model with realism for the future European electricity system in this study, and the results should not be interpreted as indicative for either system design or operational strategy for the future electricity market. The exact numbers we present in this analysis are arguably of secondary interest. More relevant are the magnitude of error in modeling results if energy system modelers use historical demand profiles or linearly scale them up as input to the model. With this study we want to deliver a message to energy system modelers regarding whether or not potential future changes in the demand pattern should be considered for modeling practice. In most of the cases investigated for this study, the altered demand patterns have relatively weak impacts on system cost, yet greater influences on the electricity supply mix are observed. Thus, if the modeler is investigating details about the future electricity supply mix or the corresponding system operation, the future electricity demand pattern needs to be taken into consideration.

One limitation for this study is that we did not consider the volume change for the electricity demand, which can be a consequence of sector coupling. Following the integration with other sectors, such as heating, transportation and industry, both the volume of the electricity demand and the demand pattern will change. In such cases, due to dramatically increased electricity demand, the impacts of different demand patterns on the modeling results might be influenced by other factors, such as the land availability constraints for VRE. Therefore, we anticipate that future studies with a good representation of sector coupling will confirm or reject the universality of some of the conclusions drawn in this paper.

**5. Conclusions and recommendations**

In this paper, we use greenfield techno-economic cost optimization models to investigate the impacts of different demand patterns on modeling results for a stylized case with three interconnected regions in Europe and one full-scale applied case (Europe). Through analyzing the system cost and electricity supply mix, we show that:

1. In most cases (zero seasonal variation, winter peak), altered seasonal demand patterns have limited impacts on system cost (<3% increase in cost compared with the current demand pattern). In contrast, a summer peak may increase the system cost by up to 8%. With additional flexible generation capacities in the electricity system, the impacts of different seasonal demand patterns become negligible;

2. The impact on system cost of a greater diurnal variation is minor (<3% increase in cost);
3. The impacts of different demand patterns on a European highly renewable electricity system are in line with the results of the stylized case, with a system cost increase up to 8% for demand patterns that have a summer peak;
4. The electricity demand pattern has a stronger influence on the electricity supply mix than on the system cost, with differences of 0%-54% for solar power and 0%-95% for storage.

In Europe, the future electricity demand pattern is uncertain, but the potential changes in demand pattern are not consequential for the system cost (with the exception of demand patterns with a summer peak). In case the future electricity demand profile shifts from the current pattern to a summer peak, using historical demand profiles to represent future electricity demand in models may result in misleading results for the system cost. Yet, a future with a summer peak in demand is indeed possible, given that there would be massive adoption of ACs to deal with the hotter summers in Europe. Since we show that such a demand pattern may have a comparatively large ( up to 8%) influence on system cost and an even larger impact (up to a factor of two) on capacity mix, it is important for modelers to exercise caution regarding the assumptions made for the future electricity demand pattern.

The conclusions that we draw in the present paper hold true for the European electricity system. They may not be valid for other regions due to, for example, different resource endowments. We anticipate future studies to test our hypotheses and to further evaluate the impact of different demand patterns on the modeling results for other regions.

**Acknowledgments**

The authors thank Niclas Mattsson for providing the GIS data. The authors also thank Emil Nyholm, Hanna Ek Fälth and David Daniels for helpful discussions and suggestions. This work was conducted as part of the ENSYSTRA project, which was supported by the European Union's Horizon 2020 research and innovation program under the Marie Skłodowska-Curie grant agreement No: 765515. L.R. and F.H. were partially funded by Chalmers Energy Area of Advance.

# Supplementary material to "*The impacts of the electricity demand pattern on electricity system cost and the electricity supply mix: a comprehensive modeling analysis for Europe*"

Xiaoming Kan[a]   Lina Reichenberg[a,b]   Fredrik Hedenus[a]

[a]Department of Space, Earth and Environment, Chalmers University of Technology, Gothenburg, Sweden

[b]Department of Mathematics and Systems Analysis, Aalto University, Helsinki, Finland

**Appendix A.**

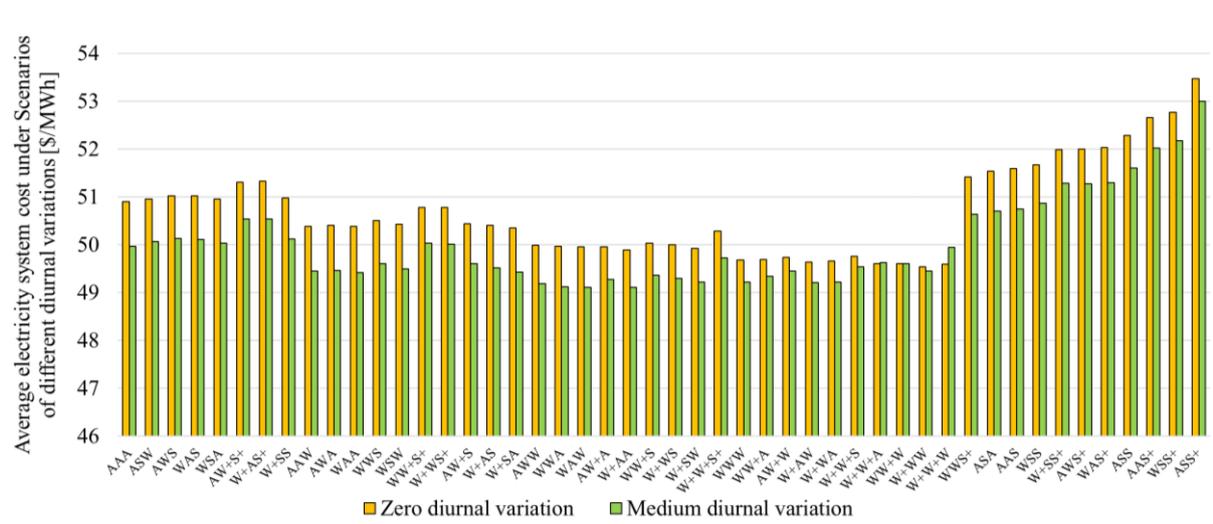

Fig. A.1. The average electricity system cost for the Scenarios with different diurnal variations in the electricity demand profile. In most cases, the Scenario of *Zero diurnal variation* has a higher system cost than the Scenario of *Medium diurnal variation*.

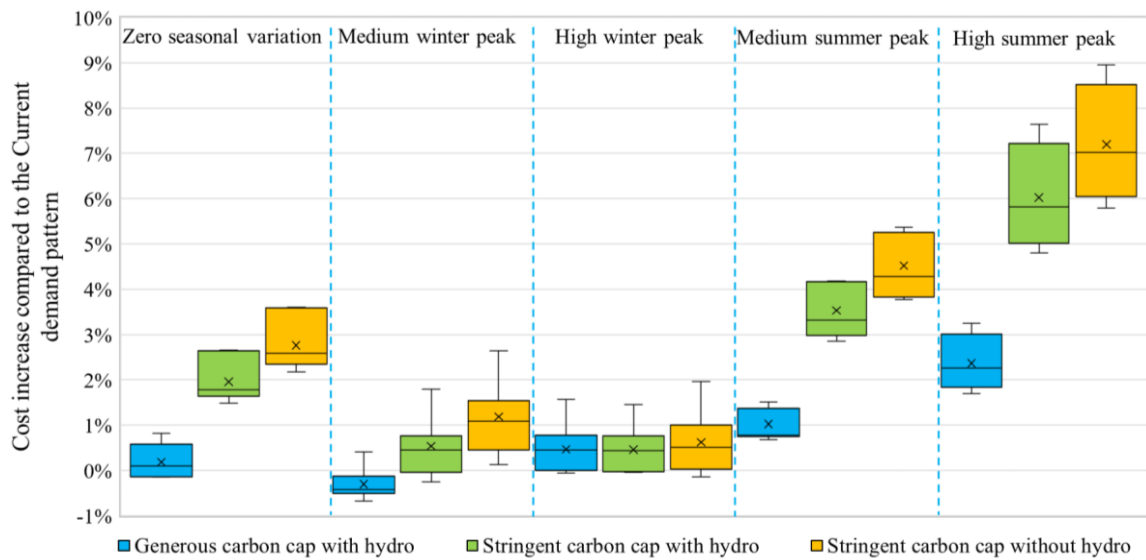

Fig. A.2. Increases in the average electricity system cost for the Scenarios of different seasonal demand patterns, as compared to the Scenario of the *Current demand pattern*, under conditions of different carbon caps and with or without hydropower. A more generous carbon cap or a larger capacity for hydropower reduces the cost increase for the Scenarios with different seasonal demand patterns compared to the Scenario of the *Current demand pattern*. The ends of the box are the upper and lower quartiles, so the box spans the interquartile range. The bar in the box represents the median value and the cross represents the average value. The whiskers are the two lines outside the box that extend to the highest and lowest values.

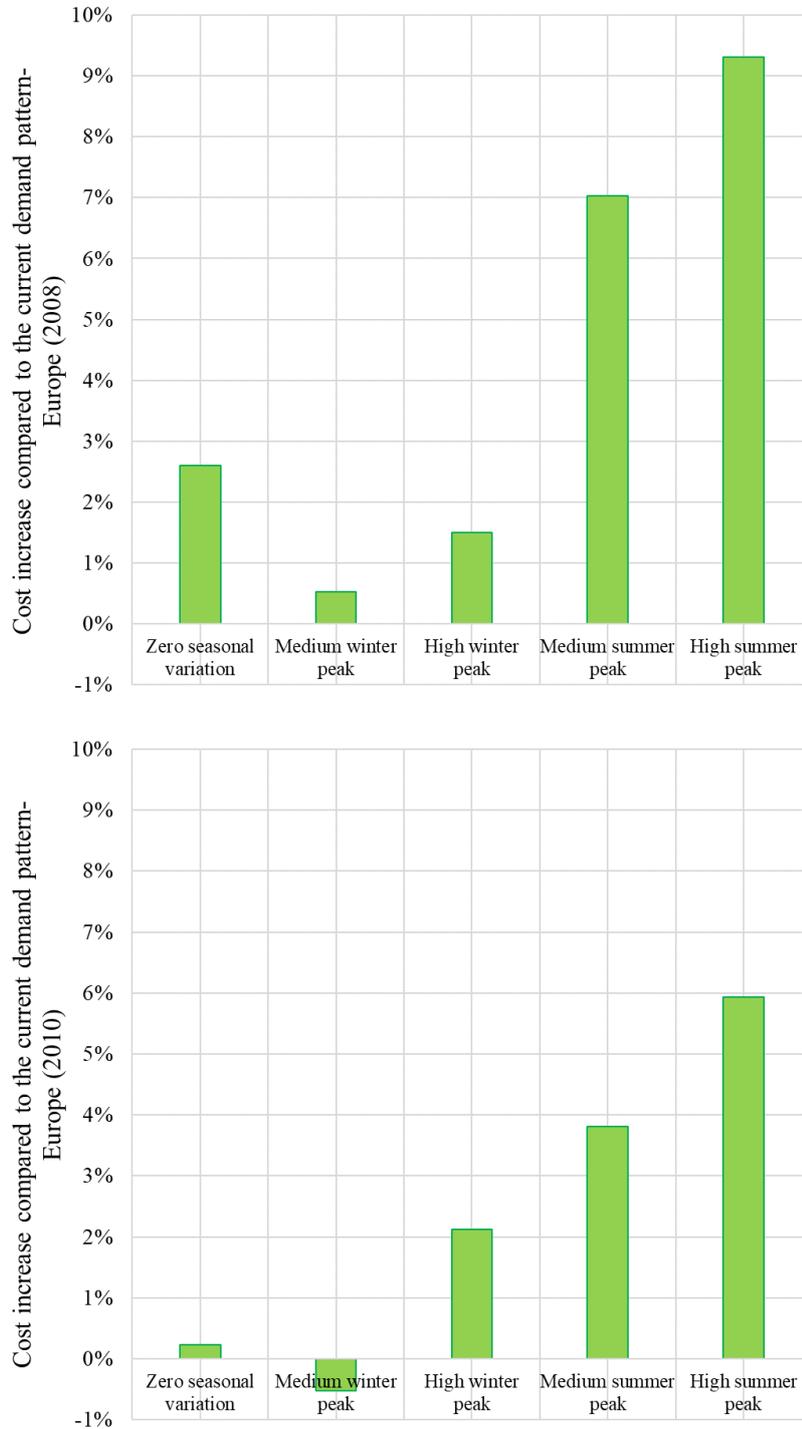

Fig. A.3. Differences in the average electricity system cost between the Scenarios of different seasonal variations and the Scenario of the *Current demand pattern* (case of Europe). The input data for wind, solar and hydropower are based on the Year 2008 (top) and 2010 (bottom) respectively.

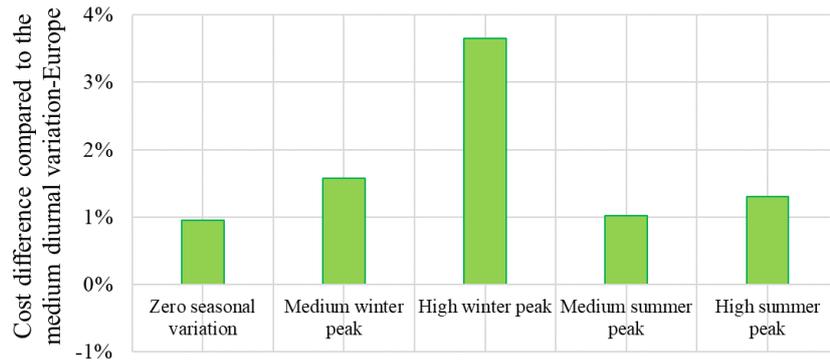

Fig. A.4. Differences in the average electricity system cost for the Scenario of *High diurnal variation* compared to the Scenario of *Medium diurnal variation* under conditions of different seasonal demand patterns for Europe. In general, the difference in system cost between the Scenarios of *Medium-* and *High diurnal variation* is limited (<4%). The input data for wind, solar and hydropower are based on the Year 2005.

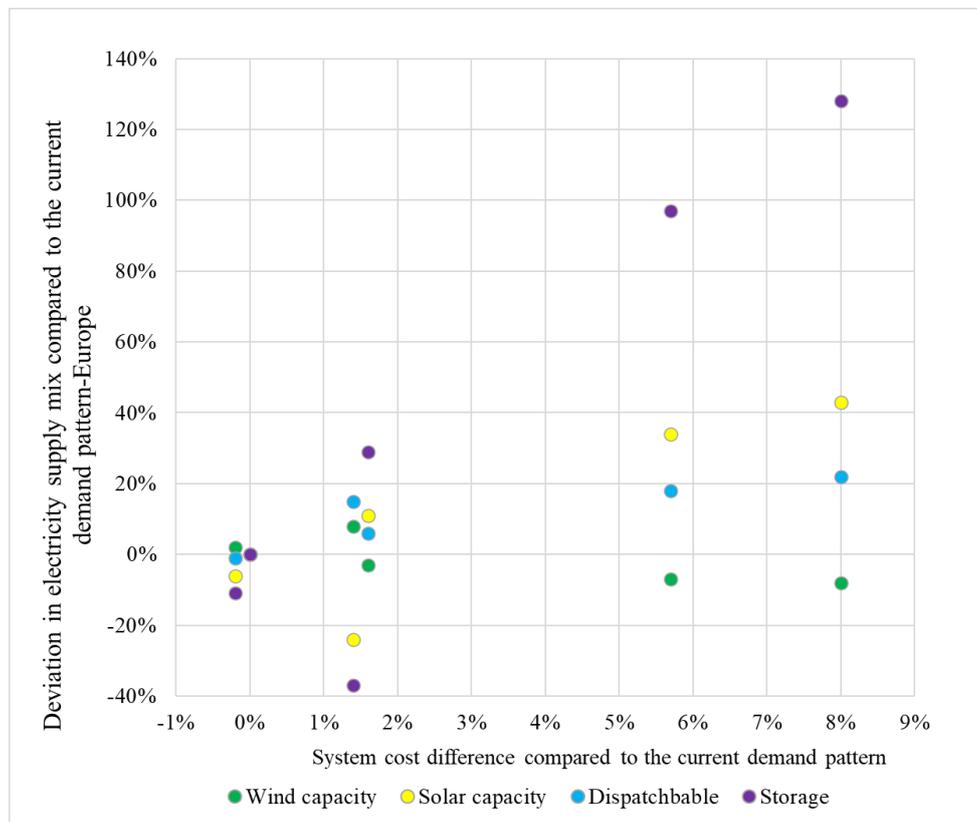

Fig. A.5. The relationship between the difference in the system cost and the deviations in the electricity supply mix for the Scenarios of different seasonal variations compared to the Scenario of the *Current demand pattern* for Europe. The input data for wind, solar and hydropower are based on the Year 2005.

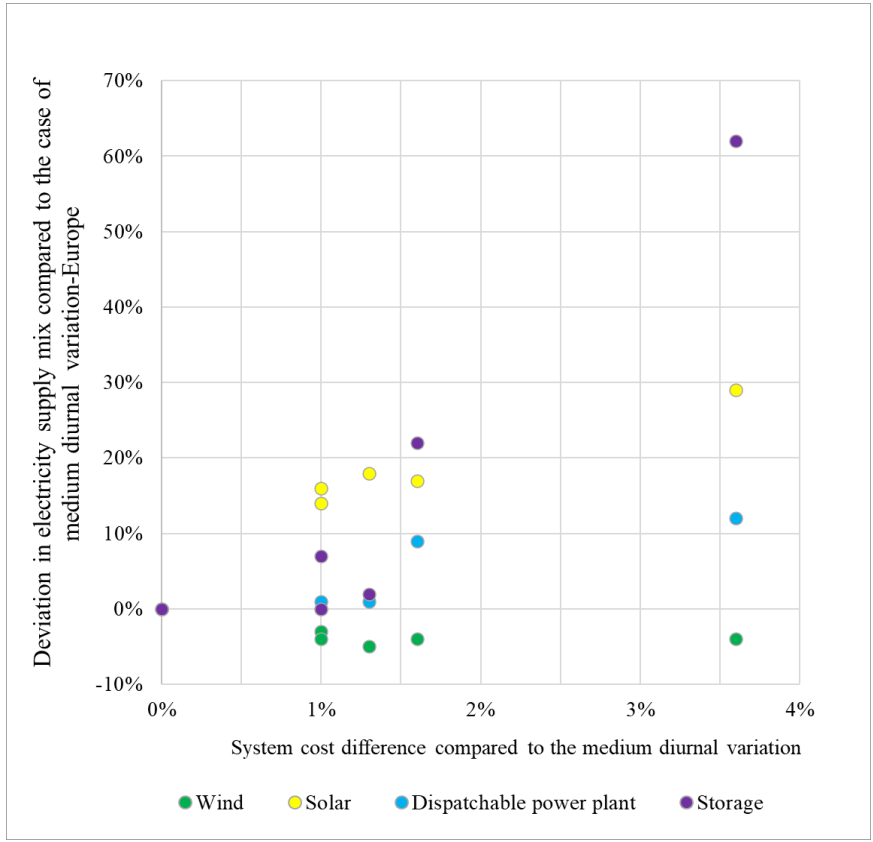

Fig. A.6. The relationship between the difference in the system cost and the deviations in the electricity supply mix for the Scenario of *High diurnal variation* compared to the Scenario of *Medium diurnal variation* for Europe. The input data for wind, solar and hydropower are based on the Year 2005.

**Appendix B.**

**The simple optimization model for the stylized case**

The mathematical representation of the simple optimization model for capacity investment and dispatch of electricity generation, transmission, storage and demand-response is described below. It is a linear optimization problem with the objective of minimizing total annual electricity system cost. It employs overnight investment in a greenfield optimization approach. Instead of investigating the transition pathway towards a low-carbon electricity system, the model seeks a cost-optimal portfolio for the future electricity system under a $CO_2$ cap.

Generation technologies included in the model are wind, solar, hydropower, biogas OCGT, biogas CCGT, natural gas OCGT and natural gas CCGT. Variation management strategies included are transmission, storage and demand-response. The model was implemented in Julia using the framework JuMP [1] and was optimized using the Gurobi solver [2]. The calculation time is around 5 minutes. A Dell Precision 5820 Tower with Intel® Core™ i9-9900X CPU @3.50 GHz, RAM 64 GB and Windows 10, 64-bit system is used for the implementation of the model. More details of the model, the objective function and the constraints are listed below:

**Sets**

| | |
|---|---|
| $r \in R$ | Regions |
| $n \in X$ | Electricity generation technologies |
| $t \in T$ | Time steps |
| $m \in M$ | Demand-response |

**Parameters**

| | |
|---|---|
| $D_{rt}$ | Demand in time-step $t$ in region $r$ $[MWh]$ |
| $C_n$ | Annualized investment cost for generation technology $n$ $[\frac{\$}{MW}]$ |
| $C^{storage}$ | Annualized investment cost for storage $[\frac{\$}{MWh}]$ |
| $C_{rr'}$ | Annualized investment cost for transmission line $rr'$ $[\frac{\$}{MW}]$ |
| $R_n$ | Variable cost for electricity generation from technology $n$ $[\frac{\$}{MWh}]$ |
| $R_m$ | Variable cost for demand-response $m$ $[\frac{\$}{MWh}]$ |
| $A_{rn}$ | Area available for VRE resource $n$ in region $r$ $[km^2]$ |
| $\rho_{rn}$ | Capacity density assigned to VRE resource $n$ in region $r$ $[\frac{MW}{km^2}]$ |
| $\eta_n$ | Efficiency for electricity generation technology $n$ |
| $\eta_\gamma$ | Efficiency for transmission |
| $\eta_s$ | Round-trip Efficiency for storage |
| $O_{rnt}$ | Output of variable quantity for VRE technology $n$ in region $r$ in time-step $t$ |
| $W_r^{max}$ | Maximum storage level of hydro reservoir in region $r$ [$MWh$] |
| $W_r^{min}$ | Minimum storage level of hydro reservoir in region $r$ [$MWh$] |
| $f_{rt}^{in}$ | Water inflow to hydropower plant in region $r$ during time-step $t$ [$MWh$] |
| $Q^{min}$ | The hourly minimum environmental flow from hydro reservoir [$MWh$] |
| $\varphi_m$ | Fraction of demand that can be curtailed in segment $m$ |
| $E_n$ | Emission factor of electricity generation technology $n$ [$\frac{gCO_2}{MWh}$] |
| $Cap$ | Carbon cap |

**Variables**

| | |
|---|---|
| $G_{rn}$ | Capacity of electricity generation technology $n$ in region $r$ [$MW$] |
| $Z_{rr'}$ | Capacity of transmission between region $r$ and $r'$ [$MW$] |
| $S_r$ | Storage in region $r$ [$MWh$] |
| $g_{rnt}$ | Electricity generation from technology $n$ in region $r$ during time-step $t$ [$MWh$] |
| $\gamma_{rr't}$ | Exported electricity from region $r$ to $r'$ during time-step $t$ [$MWh$] |
| $\gamma_{r'rt}$ | Imported electricity from region $r'$ to $r$ during time-step $t$ [$MWh$] |
| $\theta_{rt}$ | Storage level in region $r$ during time-step $t$ [$MWh$] |
| $\beta_{rt}$ | Electricity into storage in region $r$ during time-step $t$ [$MWh$] |
| $\alpha_{rt}$ | Electricity out of storage in region $r$ during time-step $t$ [$MWh$] |
| $w_{rt}$ | Hydro reservoir storage level in region $r$ during time-step $t$ [$MWh$] |
| $f_{rt}^{out}$ | Total hydro outflow from reservoirs in region $r$ during time-step $t$ [$MWh$] |
| $f_{rt}^{out'}$ | Hydro outflow through the turbine in region $r$ during time-step $t$ [$MWh$] |
| $f_{rt}^{outby}$ | Hydro outflow bypassing turbine in region $r$ during time-step $t$ [$MWh$] |
| $\lambda_{rt}$ | KKT multiplier of the demand constraint in region $r$ during time-step $t$ [$\frac{\$}{MWh}$] |
| $d_{rmt}$ | Curtailed demand in segment $m$ in region $r$ during time-step $t$ [$MWh$] |

## 1. Objective Function

Minimize total annual system cost: capacity costs for electricity generation technologies + storage costs + transmission capacity costs + electricity generation costs + demand-response costs. Therefore, the objective function is formulated as

$$Min \sum_{r,n} C_n G_{rn} + \sum_{r} C^{storage} S_r + \sum_{r,r'} 0.5 C_{rr'} Z_{rr'} + \sum_{r,n,t} R_n g_{rnt} + \sum_{r,m,t} R_m d_{rmt}. \tag{B.1}$$

## 2. Constraints

### 2.1 Demand Balance Requirement

The electricity demand has to be satisfied through generation, demand-response, trade and storage to guarantee the security of the electricity supply.

$$\sum_n g_{rnt} + \sum_m d_{rmt} + \sum_{r'} (\eta_\gamma \gamma_{r'rt} - \gamma_{rr't}) + (\eta_s \alpha_{rt} - \beta_{rt}) \geq D_{rt}, \tag{B.2}$$

where $\lambda_{rt}$ is the Karush-Kuhn-Tucker (KKT) multiplier associated with the constraint. The KKT multiplier indicates the marginal price of supplying additional demand at node *r* in hour *t* [3].

### 2.2 $CO_2$ Emission Constraints

The total $CO_2$ emissions are constrained by the carbon cap,

$$\sum_{r,n,T} \left( E_n \frac{g_{rnt}}{\eta_n} \right) \leq Cap. \tag{B.3}$$

### 2.3 Thermal Energy

The hourly thermal generation is upper bounded by capacity multiplies time interval, which is 1 *h*,

$$0 \leq g_{rnt} \leq G_{rn}. \tag{B.4}$$

### 2.4 Variable Renewable Energy

The investment in VRE capacity is constrained by area consideration and the maximum capacity density,

$$0 \leq G_{rn} \leq \rho_{rn} A_{rn}; \tag{B.5}$$

The hourly VRE generation is upper bounded by momentary weather conditions and capacity,

$$0 \leq g_{rnt} \leq O_{rnt} G_{rn}. \tag{B.6}$$

### 2.5 Hydro reservoir

For hydro reservoir, due to environmental concern, the water in the reservoir cannot exceed the maximum storage level $W_r^{max}$. In addition, the water in the reservoir is required to be above the minimum level. Therefore, the storage level has a lower bound $W_r^{min}$,

$$W_r^{min} \leq w_{rt} \leq W_r^{max}. \tag{B.7}$$

The reservoir storage level is also affected by the inflow and outflow,

$$w_{r,t+1} = w_{rt} + f_{rt}^{in} - f_{rt}^{out}, \qquad (B.8)$$

where the constraint is circular so that the storage level in the last time-step of the year equals the storage level in the first time-step of the year.

Part of the outflow, $f_{rt}^{outby}$, bypasses the turbine to balance the ecosystem and human needs for water downstream. The total outflow $f_{rt}^{out}$ has to satisfy the minimum environmental flow requirement $Q^{min}$ [4, 5]. The flow $f_{rt}^{out\prime}$ through the turbine is upper bounded by the hydro reservoir generation capacity.

$$f_{rt}^{out} = f_{rt}^{out\prime} + f_{rt}^{outby}, \qquad (B.9)$$

$$f_{rt}^{out} \geq Q^{min}, \qquad (B.10)$$

$$0 \leq f_{rt}^{out\prime} \leq G_{rn}, \qquad (B.11)$$

$$g_{rnt} = f_{rt}^{out\prime}. \qquad (B.12)$$

### 2.6 Hydro run of river

For hydro RoR, the power production is constrained by capacity and hydro inflow $f_{rt}^{in}$,

$$0 \leq g_{rnt} \leq G_{rn}; \qquad (B.13)$$

$$0 \leq g_{rnt} \leq f_{rt}^{in}. \qquad (B.14)$$

### 2.7 Transmission

The electricity traded through transmission line is upper bounded by transmission capacity,

$$0 \leq \gamma_{rr\prime t} \leq Z_{rr\prime}. \qquad (B.15)$$

### 2.8 Storage

The energy in storage is bounded by the maximum energy that can be stored,

$$0 \leq \theta_{rt} \leq S_r; \qquad (B.16)$$

The energy out from storage is bounded by the storage level in the previous time-step,

$$\alpha_{rt} \leq \theta_{r,t-1}; \qquad (B.17)$$

The energy into storage is bounded by the space left in the storage,

$$\beta_{rt} \leq S_r - \theta_{rt}; \qquad (B.18)$$

The level in storage is consistent with the charging and discharging in each hour,

$$\theta_{r,t+1} = \theta_{rt} + \beta_{rt} - \alpha_{rt}, \qquad (B.19)$$

where the constraint is circular so that the storage level in the last time-step of the year equals the storage level in the first time-step of the year.

## 2.9 Demand-side Resource

The demand-side resource adopted in this implementation is demand-response or price-responsive demand curtailment. In a given time period $t$, consumers aggregated in segment $m$ can curtail their demand at a fixed cost $C_m$. The amount of curtailed demand is upper limited by the fraction of demand in segment $m$, $\varphi_m$, times the hourly demand $D_{rt}$,

$$d_{rmt} \leq \varphi_m D_{rt}. \tag{B.20}$$

No demand rescheduling is considered in this study.

**Table B.1** Cost information for demand-response

| Cost information for demand-response | |
|---|---|
| Number of segments | 5 |
| Cost for curtailed demand in each segment[a] | 60-70-80-90-100 |
| Size of each segment [b] | 1 |
| Total price-responsive demand[c] | 5 |

[a]Cost of demand-response in each segment as a percentage of the value for the lost load ($1000/MWh)

[b]Fraction of demand-response in each segment as a percentage of hourly demand

[c]Total demand-response as a percentage of hourly demand

**Table B.2** Cost data for sensitivity analysis

| Technology | Low | Medium | High |
|---|---|---|---|
| Wind [$/kW] | 748[a] | 997 | 1246[b] |
| Solar [$/kW] | 337[c] | 674 | 1011[d] |
| Storage [$/kWh] | 78[e] | 156 | 234[f] |
| Transmission [$/MWkm] | 240[g] | 479 | 720[h] |

[a]25% reduction from medium cost level.

[b]25% increase over medium cost level.

[c]50% reduction from medium cost level.

[d]50% increase over medium cost level.

[e]50% reduction from medium cost level.

[f]50% increase over medium cost level.

[g]50% reduction from medium cost level.

[h]50% increase over medium cost level.

**Table B.3** Scenarios for the case of Europe.

| Scenario[a] | Seasonal variation | Diurnal variation |
|---|---|---|
| Current demand pattern[b] | Current pattern | Current pattern |
| Zero seasonal variation[c] | Zero | Medium/High |
| Medium winter peak[d] | 15% of annual peak demand, winter peak | Medium/High |
| High winter peak[e] | 35% of annual peak demand, winter peak | Medium/High |
| Medium summer peak[f] | 18% of annual peak demand, summer peak | Medium/High |
| High summer peak[g] | 24% of annual peak demand, summer peak | Medium/High |

[a]Similar to the stylized case, we model the combinations of seasonal and diurnal variations of the regional demand profiles for the case of Europe. For each combination (except for the Scenario of the *Current demand pattern*), all the regions have the same diurnal variation and the amplitude of the diurnal variation has two levels: *Medium* and *High*.

[b]For the Scenario of the *Current demand pattern*, all the regional demand profiles keep the current shape.

[c]For the Scenario of *Zero seasonal variation*, all the regions have demand profiles without seasonal variation.

[d]For the Scenario of *Medium winter peak*, regions in the north have a medium winter peak, regions in the central part have a medium winter peak, regions in the south have zero seasonal variation.

[e]For the Scenario of *High winter peak*, regions in the north have a high winter peak, regions in the central part have a high winter peak, regions in the south have a medium winter peak.

[f]For the Scenario of *Medium summer peak*, regions in the south have a medium summer peak, regions in the central part have a medium summer peak, regions in the north have zero seasonal variation.

[g]For the Scenario of *High summer peak*, regions in the south have a high summer peak, regions in the central parts have a medium summer peak, regions in the north have zero seasonal variation.

**Appendix C. The case of China**

We also run the REX model for China to see whether the results obtained in the stylized case is universal for regions other than Europe. In this case study, China is divided into six regions and the network topology is shown in Fig. C.1. The electricity demand for China in 2050 is estimated based on the method of [6]. The demand profiles are then treated in the same way as that for the case of Europe. The Scenarios for the case of China are presented in Table C.1.

**Table C.1** Scenarios for the case of China.

| Scenario[a] | Seasonal variation | Diurnal variation |
|---|---|---|
| Current demand pattern[b] | Current pattern | Current pattern |
| Zero seasonal variation[c] | Zero | Medium |
| Medium winter peak[d] | 15% of annual peak demand, winter peak | Medium |
| High winter peak[e] | 22% of annual peak demand, winter peak | Medium |
| Medium summer peak[f] | 19% of annual peak demand, summer peak | Medium |
| High summer peak[g] | 34% of annual peak demand, summer peak | Medium |

[a]Similar to the stylized case, we model the combinations of seasonal and diurnal variations of the regional demand profiles for the case of China. For each combination (except for the Scenario of the *Current demand pattern*), all the regions have the same diurnal variation and the amplitude of the diurnal variation is kept at the medium level.

[b]For the Scenario of the *Current demand pattern*, all the regional demand profiles keep the current shape.

[c]For the Scenario of *Zero seasonal variation*, all the regions have demand profiles without seasonal variation.

[d]For the Scenario of *Medium winter peak*, regions in the north (N, NW, NE) have a medium winter peak, SW and SE have a medium winter peak, S has zero seasonal variation. Here N refers to North China, NW refers to North West China, NE refers to North East China, SW refers to South West China, SE refers to South East China and S refers to South China.

[e]For the Scenario of *High winter peak*, regions in the north (N, NW, NE) have a high winter peak, SW and SE have a medium winter peak, S has zero seasonal variation.

[f]For the Scenario of *Medium summer peak* in China, regions in the north (N, NW, NE) have zero seasonal variation, regions in the south (S, SW, SE) have a medium summer peak.

[g]For the Scenario of *High summer peak* in China, regions in the north (N, NW, NE) have a medium summer peak, regions in the south (S, SW, SE) have a high summer peak.

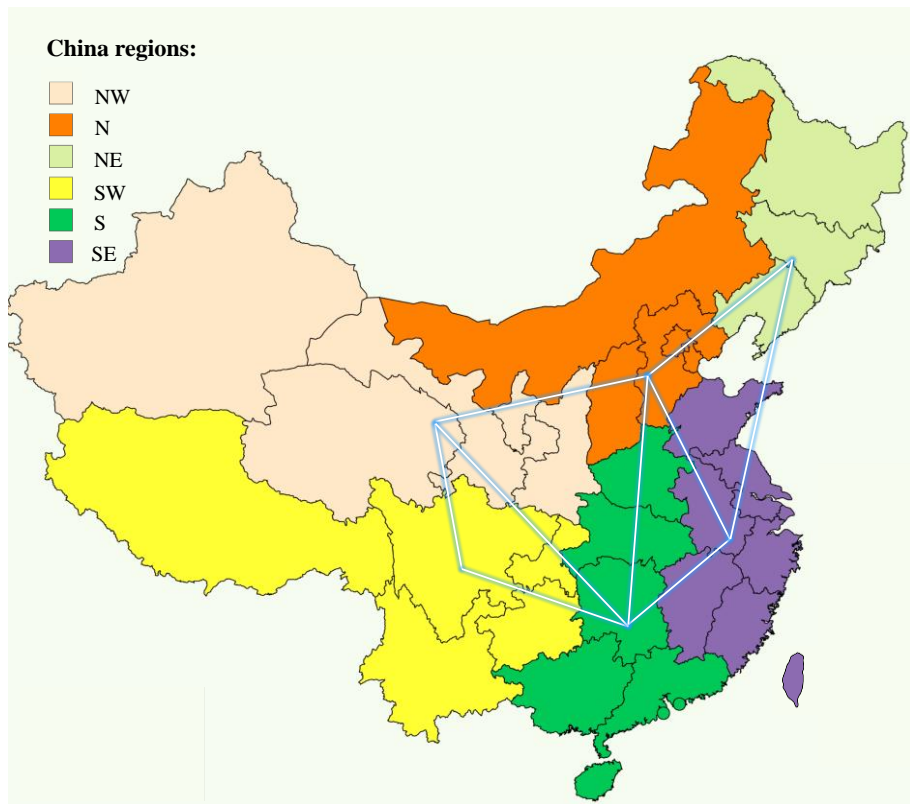

Fig. C.1. Modeled regions and the interconnected transmission networks for China.

As for the input data for the case of China, the capacities of hydro reservoir and hydro RoR are kept constant at the current level. The capacity of hydropower is taken from a previous report [7], and the hydro inflow data are based on reference [6]. The assumptions made for the biogas power plants, wind, solar, transmission, storage, demand-response, carbon cap, technology cost, and discount rate are the same as those in the stylized case. The input data for wind, solar and hydropower are based on the Year 2018 for China.

For the case of Europe, the system cost increase is less than 2% for the winter peak, while this value reaches 8% for the summer peak, as compared to the Scenario of the *Current demand pattern*. The opposite is observed for China, which shows a cost increase of up to 6% for the winter peak and cost decreases for the summer peak and zero seasonal variation, see Fig. C.2. The main reason for the discrepancy in the results for China and Europe is that hydro reservoirs in China are less capable of reserving water inflow as seasonal storage, as compared with those in Europe. Therefore, hydropower production in China is higher in the summertime when the water inflow is large, which smoothens the impact of the summer peak for China. Therefore, our results regarding the impacts of different seasonal demand patterns based on the stylized case are only valid for Europe or regions with similar resource endowments.

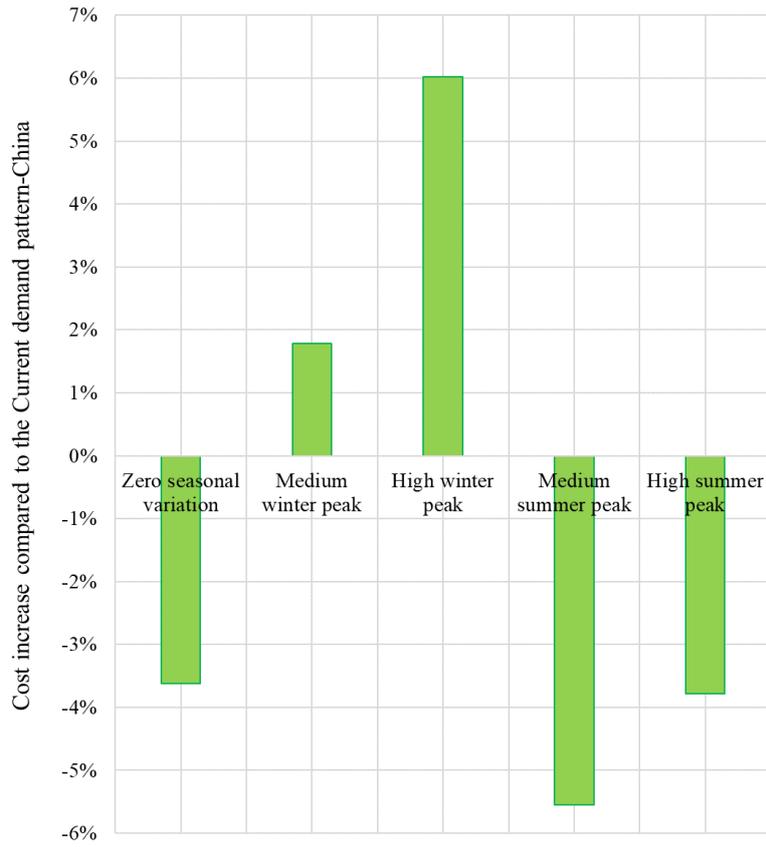

Fig. C.2. Differences in the average electricity system cost between the Scenarios of different seasonal variations and the Scenario of the *Current demand pattern* for China.